\documentclass[10pt,twocolumn]{article}

\usepackage[T1]{fontenc}
\usepackage[utf8]{inputenc}
\usepackage[english]{babel}
\usepackage{mlmodern}
\usepackage[expansion=false]{microtype}
\usepackage{amsmath,amssymb}
\usepackage{booktabs}
\usepackage{multirow}
\usepackage{array}
\usepackage{graphicx}
\usepackage{xcolor}
\usepackage{listings}
\usepackage{tikz}
\usetikzlibrary{arrows.meta,positioning,fit,backgrounds,calc}
\usepackage{hyperref}
\usepackage{url}
\usepackage{caption}
\usepackage{balance}
\usepackage[margin=0.85in]{geometry}
\usepackage{tabularx}
\usepackage{alltt}
\usepackage{textcomp}
\usepackage{fancyvrb}
\usepackage{upquote}
\usepackage{setspace}
\usepackage{fancyhdr}

\fancypagestyle{plain}{
    \fancyhf{}
    \fancyfoot[C]{\thepage}
    \fancyfoot[R]{\scriptsize[Distribution Statement A] Approved for \\public release and unlimited distribution.}
    
}

\hypersetup{
  colorlinks=false,
  linkbordercolor={1 0 0},
  citebordercolor={0 0.6 0},
  urlbordercolor={0 0.7 0.7},
  pdfborderstyle={/S/D/D [.33 .75]/W 0.75}
}

\newcommand{\qt}[1]{\textrm{``}#1\textrm{''}}

\def\ttob{\texttt{\symbol{`\{}}}
\def\ttcb{\texttt{\symbol{`\}}}}
\newcommand{\ttc}[1]{\ttob{#1}\ttcb}

\def\ttt{\texttt}
\def\trm{\textrm}

\newcommand{\itc}[1]{\textit{\it #1\/}}

\definecolor{kw}{rgb}{0.10,0.10,0.55}
\definecolor{cmt}{rgb}{0.30,0.45,0.30}
\definecolor{str}{rgb}{0.55,0.10,0.10}
\title{\bfseries Using LLMs to Adjudicate Static-Analysis Alerts with Error Reduction Techniques}

\author{%
  William Klieber\\
  \small Carnegie Mellon Univ.\\
  \small Pittsburgh, USA\\
  \small \texttt{weklieber@cert.org}
  \and
  David Svoboda\\
  \small Carnegie Mellon Univ.\\
  \small Pittsburgh, USA\\
  \small \texttt{svoboda@sei.cmu.edu}
  \and
  Lori Flynn\\
  \small Carnegie Mellon Univ.\\
  \small Pittsburgh, USA\\
  \small \texttt{lflynn@cert.org}
  \and
  Ruben Martins\\
  \small Carnegie Mellon Univ.\\
  \small Pittsburgh, USA\\
  \small \texttt{rubenm@andrew.cmu.edu}
}
\date{}

\begin{document}
\maketitle

\begin{abstract}
Static analysis is widely used for finding
security weaknesses in source code before deployment, but it often produces far
more alerts than analysts can review.
We study how well large language models (LLMs) can
\emph{adjudicate} (classify as a real bug or a false alarm) static-analysis alerts.
We use two mistake-mitigation methods:
(1) a \emph{consistency check (CC)} that runs the LLM multiple times and checks
that the verdicts are consistent with each other, and
(2) an \emph{LLM reasoning evaluation (LRE)} step that runs the LLM multiple
times and then asks the LLM to choose a verdict after evaluating the reasoning
provided by each run.

We evaluated several LLMs on three test suites: Juliet, FormAI, and SV-COMP.
Across all three suites, the mid-tier reasoning LLMs that we tested
(o4-mini, gpt-oss-120b, gpt-oss-20b) reach high
recall (percent of real bugs that the tool correctly flags as needing repair /
manual attention) and specificity (percent of actually false alerts that the
tool correctly dismisses as false alarms).  With mistake mitigation, they reach
at least 98\% recall and at least 94.8\% specificity on every suite
(with CC alone on Juliet and SV-COMP, and with LRE+CC on FormAI).

We probe Juliet memorization and show that o4-mini can often
reconstruct sanitized test cases' original identities, so we base our
generalization claims primarily on FormAI, scored against our own unpublished manual
adjudications.
A complementary \emph{flipped-verdict} experiment suggests that o4-mini does
exercise its reasoning capabilities on Juliet rather than reciting a memorized
verdict, but doesn't definitively rule out the possibility of overfitting.
We also note a few cases where the LLM disagreed with our initial manual
adjudications but the LLM's explanation of its answer convinced us that its
answer was correct and our initial manual adjudication was wrong.
We also report results of using the LLM to synthesize a program that
dynamically \emph{triggers} the flaw
as independent evidence; a
validity check rejected every trigger driver aimed at a false alarm, so a
valid trigger proved to be strong evidence of a real flaw.
\end{abstract}

\section{Introduction}
\label{sec:intro}

It is a standard step in software development to evaluate source
code for security weaknesses before it is fielded.
Static analysis (SA) is widely used and is among the
best automated techniques available, but using it well requires substantial
manual effort: a tool will typically report many alerts, a large fraction of
which are false positives, and an analyst must \emph{adjudicate} each
alert (decide whether it indicates a real flaw or not).
The volume of alerts is often too large to
review in its entirety, so teams triage. A common practice is to
manually adjudicate only the highest-severity alerts and to leave the remainder unreviewed.
Unreviewed alerts constitute \emph{unknown risk}: a real vulnerability may
be hiding among them.

Recent large language models (LLMs) change what is feasible.
Unlike earlier machine-learning approaches, modern reasoning LLMs
produce a detailed chain of reasoning leading to their conclusion, and that
reasoning can be double-checked. LLMs can also request
information they lack (e.g., the definition of a struct or macro) and a
driver program can retrieve and supply it. Several groups have begun to
apply LLMs to static-analysis alerts and to false-positive
reduction~\cite{li2023assisting,wen2024llm4sa,du2026tencent}.

This paper studies how well modern LLMs can perform this adjudication. We built
{LASAA} (\emph{LLMs for Adjudication of Static-Analysis Alerts}), an
open-source, analyzer-agnostic pipeline that adjudicates each alert with an LLM and
reports a justification with every verdict, and we use it as the instrument
for an empirical study on three benchmark test suites. Our aim is to adjudicate a
large fraction of alerts automatically with high accuracy, so that analysts
can focus their limited attention on the alerts that genuinely require it.

\paragraph{Contributions.}
\begin{enumerate}\itemsep2pt
\item An \emph{LLM reasoning evaluation} (LRE) step that asks the LLM
  to reconcile its own discordant runs by weighing their reasoning
  (Sec.~\ref{sec:approach}), together with an evaluation showing that
  combining LRE with a consistency check (CC) sharply reduces \ttt{uncertain}
  verdicts compared to CC alone (without LRE).
  With reasoning models and the right CC threshold, LRE+CC matches or exceeds a
  plain majority-vote baseline for both recall and specificity on our \hbox{FormAI} benchmark.
\item An empirical study of LLM alert adjudication on Juliet, FormAI, and
  SV-COMP that reports the rates of two types of adjudication errors (missed flaws
  and false alarms), finding that reasoning LLMs achieve high
  recall and specificity, even the small open-weight gpt-oss-20b when paired
  with mistake mitigation.
\item A direct measurement of Juliet memorization via a
  filename-reconstruction probe, showing that o4-mini can often recover a
  sanitized case's identity, together with a \emph{flipped-verdict} experiment
  that provides some evidence that o4-mini nonetheless adjudicates from the code rather than from a
  memorized verdict.
\item A \emph{dynamic trigger test} that seeks execution-based evidence for an
  alert adjudicated as a true positive, paired with an automated validity
  check that rejects trigger drivers reaching the flaw only by violating the
  program's preconditions (Sec.~\ref{sec:trigger}).
\item An analysis of how hard it is to obtain trustworthy ground truth,
  including cases where LLM output or the trigger test convinced us that we
  needed to correct our initial manual adjudications
  (Sec.~\ref{sec:groundtruth}).
\item The LASAA tool released as an artifact:\\ \url{https://github.com/cmu-sei/lasaa/}.
\end{enumerate}

\section{Background and Problem}
\label{sec:background}

\paragraph{Alert adjudication.}
An alert from a static analyzer includes the location (filename and line number) and
the weakness type (often with a CWE identifier~\cite{cwe}). The \emph{ground truth}
for an alert is whether the indicated weakness is actually present at the indicated
location.

\paragraph{Two error types.}
It is useful to view an LLM-based adjudicator as a binary classifier whose
positive class is ``a real flaw is present''.
The positive class is subdivided into three subclasses:
(1) Strong positive,
(2) Dependent (see below), and
(3) Weak positive (``\ttt{uncertain}'').
There are two distinct ways for the adjudicator to be wrong, with very
different costs:
\begin{itemize}\itemsep1pt
\item \textbf{Missed flaw (adjudicator false negative):} the alert is actually
  true (a real bug), but the adjudicator says ``false''. In a high-assurance
  setting, this is the dangerous error: a real vulnerability is cleared and
  may be fielded.
\item \textbf{False alarm (adjudicator false positive):} the alert is actually
  a false alarm, but the adjudicator says ``true''. This is comparatively
  safe, but an excessive number of false alarms creates too great a burden
  for analysts/developers.
\end{itemize}
Throughout our evaluation (Sec.~\ref{sec:eval}) we report
\emph{class-conditional} metrics that capture the rates of these two types of
adjudication errors:
\begin{itemize}
\item recall (percent of real bugs that the tool correctly flags as needing repair /
manual attention), and
\item specificity (percent of actually false alerts that the
tool correctly dismisses as false alarms).
\end{itemize}

\paragraph{Dependent alerts.}
We use the term \emph{dependent alert} for the case in which fixing an
earlier-executed line (with the same flaw type) also fixes the current
line~\cite{svoboda2016static}.  For example, consider the C code below:
\begin{samepage}
\begin{verbatim}
  struct Foo *x = malloc(sizeof(struct Foo));
  x->field1 = 1;
  x->field2 = 2;
\end{verbatim}
\end{samepage}
Here, the assignment to \ttt{x->field2} has a flaw (because \ttt{x} isn't checked for
being NULL), but this null-pointer dereference would have already been tripped
in the assignment to \ttt{x->field1}, so we mark the assignment to \ttt{x->field2}
as dependent.  For dependent alerts, we ask the LLM to cite the line that it
depends on and argue that the alert disappears once that line is fixed.
The ``dependent'' adjudication type is useful because pointing a developer to the line
that actually needs fixing is more useful than flagging every downstream symptom.
Also, by asking the LLM to identify which alerts are dependent, we exercise
more of the LLM's reasoning capabilities than we would with only a binary true/false call.

\section{Approach}
\label{sec:approach}

\begin{figure*}[t]
\centering
\includegraphics[width=0.75\textwidth]{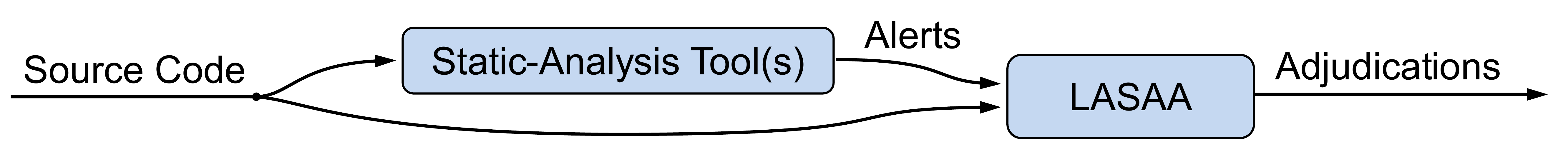}
\caption{LASAA pipeline}
\label{fig:arch}
\end{figure*}

\subsection{Overview}
Figure~\ref{fig:arch} shows an overview of the LASAA pipeline. One or more
static analyzers run over the source code and produce alerts. For each alert,
LASAA uses an LLM to adjudicate the alert and reports both the final verdict
and reasoning to support the verdict.

\subsection{Building the query}
In the simplest case, LASAA issues one query per alert. The query contains
(A)~the alert fields (file, line, CWE(s), alert message, etc.),
(B)~the source code of the function that contains the flagged line, with
\hbox{``\texttt{// Line N}''} annotations appended, and
(C)~instructions on what to do.

The LLM is instructed to ask for definitions of macros and structs that it needs.
If the LLM makes such a request, LASAA locates the definition of the
macro/struct (using
\ttt{ctags}\footnote{\url{https://github.com/universal-ctags/ctags}}) and
re-runs the prompt with the definition appended.
(This functionality was not exercised by the benchmarks in the Evaluation
section, because the programs were small enough that they easily fit in
the LLM's context window.  However, in ad-hoc testing on \ttt{dos2unix}
(a real-world codebase, with 3,900 lines of code), the LLM
did ask for definitions.)

\paragraph{Prompt design.}
The instructions ask the LLM to classify the alert as \ttt{true},
\ttt{false}, \ttt{dependent}, or \ttt{uncertain}.%
\footnote{Although ``\ttt{uncertain}'' was offered as an option, it was never
selected by an LLM in our experiments.}
For alerts classified as true, we ask the LLM to give a trace demonstrating the
vulnerability.
For alerts classified as false, we ask the LLM to give a proof sketch arguing
why it is a false positive.
The LLM is instructed to include its final answer in JSON form (e.g.,
\ttt{\ttc{"verdict":\ "false"}}) at the end of its response.

\subsection{Mitigating LLM mistakes}
\label{sec:MMM}
LASAA implements two complementary mechanisms that are also independently selectable:
a consistency check (CC) and an LLM reasoning evaluation (LRE).

\begin{figure}[t]
\centering
\includegraphics[width=\columnwidth]{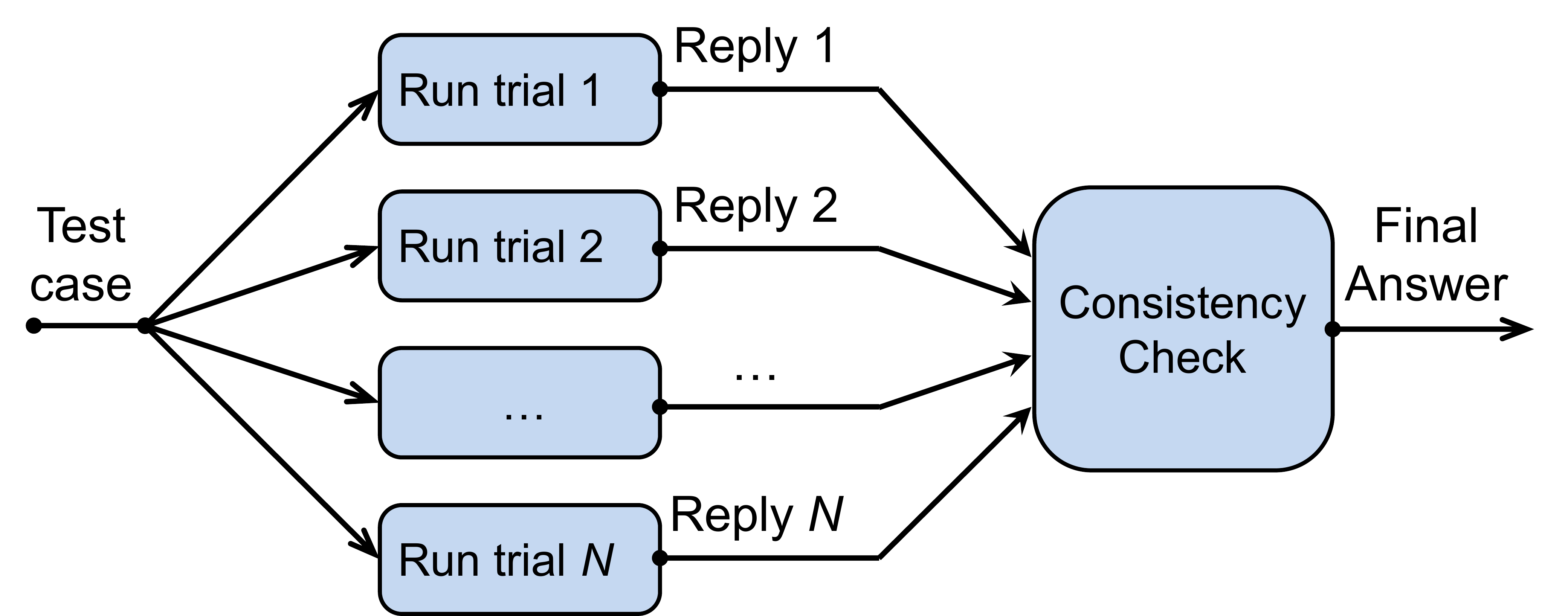}
\caption{Consistency check.}
\label{fig:consistency}
\end{figure}

\subsubsection{Consistency Check (CC).}
The tool runs $N$ independent trials of the query (default $N{=}10$). If a
single verdict is reached on at least a \hbox{threshold} percentage of
trials, that verdict is returned; otherwise the tool returns
\qt{\ttt{uncertain}}, leaving the alert for manual attention
(Fig.~\ref{fig:consistency}).
A higher threshold yields fewer \emph{wrong} answers but at the cost of more
\ttt{uncertain} answers.
In our initial experiments we chose, per
LLM/suite, the smallest threshold that drives the percentage of wrong answers
below~5\%.
For the revised experiments, we decided on a uniform threshold of 80\% for all
LLMs and suites before running them.
This choice was somewhat arbitrary but
informed by the results of the initial round.
For the non-frontier reasoning LLMs that we tested with FormAI, we also
show results for a threshold of 70\%.

In most of our experiments, we ran 10 trials and used a threshold of 80\%.
With these settings, an outcome of \ttt{uncertain} arose only if at least
3 of the 10 trials reached a verdict in \{\ttt{true}, \ttt{dependent}\}.
(The LLM itself never returned a verdict of \ttt{uncertain} in our experiments.)
Since false negatives are significantly more costly than false positives,
we treat \ttt{uncertain} as a weak positive.

If the consistency check fails, LASAA can optionally ask the LLM to briefly
explain the source of the disagreement, which may be useful for manual review.

\subsubsection{Majority-vote baseline.}
As a baseline, LASAA also supports plain \emph{majority voting} instead of the
above-defined consistency check.
This is a natural alternative to our consistency check: it
returns the most common verdict and never returns \qt{\ttt{uncertain}}.
Because our verdicts are not binary, we take the majority in two stages:
\begin{enumerate}
\item
    Negative (\ttt{false}) vs.\ positive (\ttt{true}/\ttt{dependent}), with
    ties being resolved in favor of positive.
\item
    If positive, then \ttt{true} vs.\ \ttt{dependent}, with ties being resolved
    in favor of \ttt{true}.
\end{enumerate}
Unlike the consistency check, majority voting never returns \ttt{uncertain} and
offers no knob to trade missed flaws against false alarms.
In the data tables, a value of \qt{\ttt{maj}} in the ``CC'' column indicates
the majority-vote baseline.

\begin{figure*}[t]
\centering
\includegraphics[width=\textwidth]{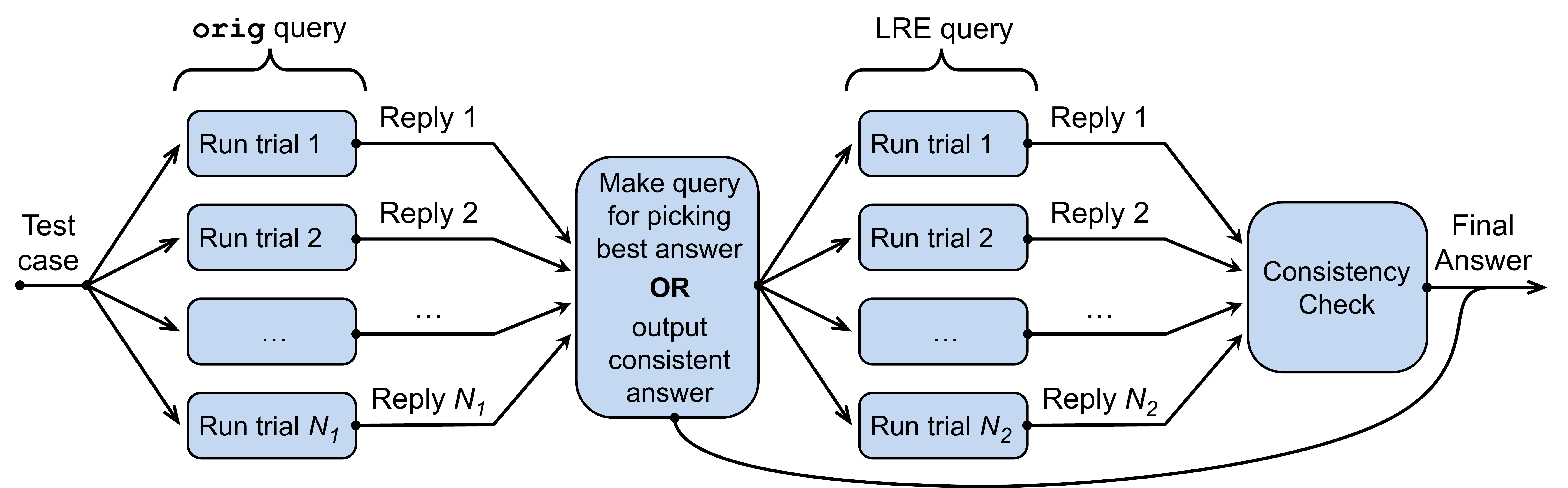}
\caption{Combining LLM reasoning evaluation with consistency check}
\label{fig:lre-and-cc}
\end{figure*}

\newcommand{\centh}[1]{\multicolumn{1}{c|}{#1}}

\begin{table*}[t]
\renewcommand{\arraystretch}{1.3}
\centering\small
\begin{tabular}{|l|r|r|r|r|r|r|r|r|r|r|r|}
\hline
Repl. &\centh{1}&\centh{2} &\centh{3} &\centh{4} &\centh{5} &\centh{6} &\centh{7} &\centh{8} &\centh{9} &\centh{10} & TOTAL \\
\hline
Orig &  5/10 &  2/10 &  1/10 &  1/10 &  1/10 &  2/10 &  3/10 &  1/10 &  0/10 &  3/10  & 19/100 \\
\hline
LRE  & 10/10 & 10/10 & 10/10 & 10/10 & 10/10 & 10/10 & 10/10 &  9/10 &  \centh{---}  & 10/10  & 89/90 \\
\hline
\end{tabular}
\caption{The car-wash problem: number of correct (``drive'') verdicts for original and LRE queries.}
\label{tab:car-wash}
\end{table*}

\subsubsection{LLM Reasoning Evaluation (LRE).}
When trials disagree, instead of returning ``\ttt{uncertain}'',
LASAA can present the original question and the discordant responses back to
the LLM and ask it to evaluate the competing reasoning.
The easiest way to think about LRE, and the shortest way to describe it,
is that we ask the LLM to ``pick the best answer'' (as in LLM-as-a-judge).
This is also the language we use in our diagrams of the LRE process.
The actual prompt, however, is somewhat different:

\begin{quote}\small
Below is a question and discordant responses to it.  Carefully evaluate these
responses.  Then, write your own response to the original question.  Your
response should also briefly indicate what you find wrong/unconvincing about
responses that reached a different final answer.  (You don't need to address
each input response individually; just briefly point out what the flaws are.)
\end{quote}

The motivating idea behind LRE is that, when generating a response, the LLM sometimes misses
a decisive consideration that it would readily recognize as important once raised.
Running $N$ trials raises the chance that at least one response surfaces such a
consideration, so that the LLM has it in mind when evaluating the reasoning of
the trials.
%
We tested this premise on the \itc{car-wash problem}~\cite{carwash-origin,carwash-viral},
a prompt that circulated online as a test of LLM informal reasoning and that
others have also used to evaluate LLMs~\cite{carwash-evals}.
We used the following prompt:

\begin{quote}\tt\small
    I need to get my car washed. The car wash is
    50 meters away. Should I drive there or walk?
    At the *end* of response, say
    \`{}\ttc{"verdict":\ "<ANSWER>"}\`{} where <ANSWER> is
    your final answer (either "drive" or "walk").
\end{quote}

We ran 10 replications of 10 trials each, using \ttt{o4-mini}.
In one replication, all 10 trials unanimously
agreed on \qt{walk} as the verdict, so LRE was skipped for that replication.
In the other 9 replications, we ran LRE 10 times.
The results for each replication are shown in Table~\ref{tab:car-wash}.
Of the 90 total LRE trials, 89 correctly gave a verdict of \qt{drive} and only one
gave an incorrect verdict of \qt{walk}.
In contrast, in the 100 trials of the original prompt, 81 gave an incorrect
verdict of \qt{walk}, and only 19 gave the correct verdict of \qt{drive}.

\begin{figure}[tb]
{
\small
\fontsize{8.75}{11}\selectfont
\begin{Verbatim}[frame=single]
func adjudicate(alert, thres, use_LRE, N1, N2):
    orig_query = build_orig_query(alert)
    orig_replies = ask_LLM(orig_query, N1)

    if use_LRE and not is_unanimous(orig_replies):
        lre_query = build_LRE_query(orig_query,
                                    orig_replies)
        final_replies = ask_LLM(lre_query, N2)
    else:
        final_replies = orig_replies

    return consistency_check(final_replies, thres)

func consistency_check(replies, thres):
    v = most_frequent_verdict(replies)
    if fraction_with_verdict(replies, v) >= thres:
        return pick_reply_with_verdict(replies, v)
    else:
        return '{"verdict": "uncertain"}'
\end{Verbatim}
}
\vspace{-2ex}
\caption{Pseudocode for CC and LRE, where \ttt{N1} and \ttt{N2} are the number
of times to run the \ttt{orig} query and the LRE query, respectively;
see Sec.~\ref{pipeline-phases-and-caching} for details.
The function \ttt{ask\_LLM(\itc{query}, \itc{N})} sends \itc{query} to the LLM
\itc{N} times and returns the $N$ replies from the LLM.
}
\label{fig:pseudocode-cc-lre}
\end{figure}

A consistency check can be applied to the LRE prompt, as shown in Fig.~\ref{fig:lre-and-cc}:
LASAA runs the LRE prompt $N$ times and returns ``\ttt{uncertain}'' unless the
results are consistent on a given percentage of runs.
Fig.~\ref{fig:pseudocode-cc-lre} gives the combined control flow (omitting
details such as on-disk caching, struct/macro lookup, and the option for the
majority-vote baseline).
With LRE enabled, there are two types of queries where we ask the LLM to give a verdict:
(1) the original query asking the LLM to adjudicate the alert, and
(2) the LRE query asking the LLM to evaluate the replies to the original query.
We use the term ``\ttt{orig} query'' to refer to the first type of query.

\subsubsection{Pipeline and caching}
\label{pipeline-phases-and-caching}
LASAA's adjudication pipeline has two phases:

\begin{itemize}
\item Phase 1:
    Run the original prompt $N_1$ times.
    When LRE is on, these runs supply the responses that the LRE step evaluates.
    When LRE is off, they are the consistency-check sample.
\item Phase 2 (run only when LRE is on and the phase-1 verdicts are not unanimous):
    Run the LRE prompt $N_2$ times, presenting the phase-1 responses for evaluation.
\end{itemize}

The final verdict comes from the last phase that ran: with LRE off, the
consistency check (CC) is applied to the $N_1$ phase-1 verdicts; with LRE on,
it is applied to the $N_2$ phase-2 verdicts (unless phase~1 was unanimous, in
which case the unanimous verdict is returned and phase~2 is skipped).
For the \qt{\ttt{CC=maj}} rows in the data tables, the majority-vote algorithm
was substituted for the CC algorithm.

When both \ttt{CC=no} (neither CC nor majority-vote) and \ttt{LRE=no}, we use
$N_1 = 1$.
When \ttt{LRE=yes} but \ttt{CC=no}, LASAA runs multiple phase-1 trials
(for LRE input) but only a single LRE trial ($N_2 = 1$).
When both CC and LRE enabled, we used $N_2 = N_1$.

Intermediate queries and replies are cached on disk so that re-running with
different options reuses prior LLM calls when possible.
For example, running with
(\ttt{CC=no}, \ttt{LRE=yes}) issues the original query ten times, and a subsequent run with
(\ttt{CC=80\%}, \ttt{LRE=no}) reuses those ten trials. This caching also makes our
CC-only vs.\ CC+LRE comparisons more apples-to-apples, since the phase-1 trials
are shared.

\paragraph{Relation to USC.}
LRE is closely related to Universal Self-Consistency (USC)~\cite{chen2023-USC},
which likewise generates multiple responses to a question using an LLM and then
feeds those responses back to the LLM to get a final response.
The key difference is in what the LLM is asked to do: USC asks it to
``Select the most consistent response based on majority consensus'',
whereas LRE asks it to weigh the competing reasoning.  This distinction is
illustrated by the car-wash experiment above: LRE picks the minority position
because it is better reasoned than the majority consensus.
AgentAuditor~\cite{AgentAuditor-Feb2026} makes the same observation that drives
LRE: that a simple majority vote ignores reasoning and lets flawed reasoning of
a majority override a correct minority.

\paragraph{Relation to LLM-as-a-judge.}
LRE also invites comparison to the \emph{LLM-as-a-judge}
paradigm~\cite{zheng2023judging,gu2024judgesurvey}, in which an LLM scores or
ranks candidate responses as a scalable stand-in for human evaluation.
LRE shares the paradigm's central premise (that an LLM can effectively
evaluate responses) but differs in some aspects.

In early and widely cited LLM-as-a-judge settings such as MT-Bench and Chatbot Arena, the
task is often open-ended (e.g., judging chatbot response quality) without
objective ground truth.
More recent work also uses LLM judges for inference-time adjudication on
verifiable reasoning tasks.\cite{AgentAuditor-Feb2026}
In LRE, the actual wording of the prompt directs the LLM to produce a new response
to the original question.
The LLM might adopt a candidate answer but is free to depart
from all of them.
LRE is thus better described as an accuracy-improvement
technique that uses judging as its mechanism, in the same family as
self-consistency~\cite{wang2022self} and USC~\cite{chen2023-USC}, than as an
instance of LLM-as-a-judge evaluation.
Known weaknesses of LLM judges nonetheless remain relevant: position and
verbosity biases~\cite{zheng2023judging} could affect which reasoning the LRE
step finds convincing.
Self-preference bias~\cite{panickssery2024selfpref} is not relevant to the
experiments we have conducted so far (since all input responses are authored by
the same LLM), but it would come into play if multiple different LLMs were
used to generate the input responses.

\section{Implementation Details}
\label{sec:impl}

\paragraph{Alert ingestion.}
LASAA is static-analyzer-agnostic: it ingests alerts in a small common format,
with converters for SARIF\footnote{\url{https://docs.oasis-open.org/sarif/sarif/v2.1.0/sarif-v2.1.0.html}} and other formats.

\paragraph{Locating the enclosing function.}
To locate the function that contains a flagged line, LASAA runs \texttt{ctags}
over the project and records the starting line number and ending line number of
every function.
A binary search over those ranges maps an alert's line number to its function.
The function body, not the whole file, is the default context, which keeps
queries small while preserving the locally relevant code.

\paragraph{On-demand definitions.}
The flagged function often references symbols defined elsewhere (structs,
macros, helper functions). The prompt allows the LLM to emit a line of the
form \verb|{"need_defs": [...]}| listing symbols it needs. When LASAA sees such
a request, it looks the symbols up (again via the \texttt{ctags} database),
appends their definitions to the prompt, and re-issues the query. This loop
repeats until the LLM produces a verdict or a maximum number of attempts is
reached.

\section{Dynamic Trigger Test}
\label{sec:trigger}

A verdict of \emph{true} indicates that a real flaw exists.
We want external, execution-based evidence for that claim.
LASAA's \emph{trigger test} gathers such evidence, by asking the LLM to
synthesize a driver that makes the flaw manifest at run time under a
debugger and suitable instrumentation, such as UBSan~\cite{sanitizers}.

If the LLM finds this task difficult or paradoxical, it might generate an invalid driver
that manifests the flaw by ``cheating''.
An example of cheating is passing a \texttt{NULL} pointer to a function
that is documented as not gracefully handling \texttt{NULL}.
Only a driver that both triggers the alert \emph{at the flagged line} and does so without cheating is
accepted as evidence of a true positive.

\paragraph{Synthesizing and running a trigger.}
The first query gives the LLM the alert and the line-annotated source and
asks it to write a driver that sets up any data structures and then invokes the vulnerable function
(or one of its callers)
so that the flagged line should behave as the alert describes.
When synthesizing the driver, the LLM is instructed not to cheat;
it must respect any standard function's contract.
For example \texttt{malloc} must return a pointer to valid memory or \texttt{NULL}.
The LLM may also decline, returning a plain-text explanation and no code,
if it cannot construct a trigger driver.

The original source is compiled with both AddressSanitizer and
UndefinedBehaviorSanitizer and with \texttt{main} renamed
and the driver is linked with it.
The \verb|-Dmain=main_orig| rename avoids any potential clash with the driver's own \texttt{main}.

\begin{lstlisting}[basicstyle=\ttfamily]
clang -c -g -O0 -Dmain=main_orig \
  -fsanitize=address -fsanitize=undefined \
  -fno-omit-frame-pointer source.c
clang -c -g -O0 -fsanitize=address \
  -fno-omit-frame-pointer trigger.c
clang -g -O0 -fsanitize=address \
  trigger.o source.o -o trigger
\end{lstlisting}

The linked program is then run under \texttt{gdb} with a breakpoint
at the flagged line; execution stops there, emits a marker, and single-steps over the line.
The test \emph{succeeds} only if a sanitizer- or signal-detected error is produced.
These can include a segmentation fault, a floating-point exception,
a signed-overflow report, or an AddressSanitizer error
\emph{at} the flagged line.
Any such error that occurs \emph{before} the breakpoint is considered a trigger failure.
The evidence must implicate the flagged line itself, not some earlier line in the run.
The LLM was given three chances to produce a working driver;
a fourth chance yielded no additional successes in our experiments.

The LeakSanitizer instrument does not work well under GDB;%
\footnote{\url{https://stackoverflow.com/questions/54022889}}
consequently this technique does not correctly detect memory leaks,
so we excluded the three memory-leak alerts from the trigger-test evaluation.
In theory, we could use Valgrind to detect memory leaks; this is future work.

\paragraph{Validity check.}
A successful driver is only meaningful if it respects the program's preconditions:
that is, it does not ``cheat''.
A driver that cheats by passing obviously invalid arguments to the entry point,
or that stubs a library function in a way that violates contracts,
can manufacture a ``vulnerability'' that no realistic caller could reproduce.
Our attempts to prevent cheating by instructing the LLM not to cheat have been unsuccessful,
and we have found that a better approach is to ask the LLM, via a separate query,
to confirm that the driver does not cheat.
For example, if the driver calls \texttt{main\_orig},
it must provide \texttt{main\_orig} with a well-formed \texttt{argv}
null-terminated list of null-terminated strings.
Likewise, the driver is allowed to stub any standard C or POSIX functions, such as \texttt{malloc},
but if the driver does so, then any stubbed function must respect its contract.
For example, \texttt{malloc} must return a pointer to valid memory or \texttt{NULL}.
Also, any pointer passed to a function must be valid and dereferenceable.
The LLM must respond to the query with true or false with a rationale.
A driver that triggers the alert but fails this check is \emph{invalid} and is discarded;
a driver that triggers the alert \emph{and} passes is a \emph{valid trigger} and counts as
execution-based evidence of a true positive.

\section{Evaluation}
\label{sec:eval}

We ran each benchmark suite in two rounds. The \emph{initial} round used our
initial prompts and answer keys.  Analyzing its errors exposed concrete,
fixable problems: ambiguous prompt instructions, a few answer-key mistakes, and
a mismatch between our prompt and SV-COMP's conventions about \texttt{malloc}
and stack-allocated variable-length arrays. We corrected these and re-ran a
\emph{revised} round.  For each benchmark suite, we first present the results
of the initial round followed by the results of the revised round, so that the
effect of the changes is visible. Our discussion emphasizes the revised
results, since the initial round's flaws are now understood. The tables omit
the date suffix from the names of the closed LLMs; the dated versions are:
\texttt{gpt-4o-2024-08-06},
\texttt{o4-mini-2025-04-16},
\texttt{gpt-5.4-2026-03-05}, and
\texttt{gpt-5.5-2026-04-23}.
We also test the open-weight LLMs \texttt{gpt-oss-20b} and \texttt{gpt-oss-120b}.

All queries were issued with each provider's default sampling settings; we
did not specify a temperature\footnote{%
The closed reasoning models we tested (o4-mini, gpt-5.4, gpt-5.5, opus-4.8) do
not even support setting the temperature.}
or other such parameters.  For the LLMs \texttt{gpt-5.4}, \ttt{gpt-5.5}, and
\ttt{opus-4.8}, we set the reasoning effort to \ttt{high};
we inadvertently left the reasoning effort at the default setting for \ttt{o4-mini}, and
the APIs of the other LLMs we tested offer no such setting.
The prompt templates are included in the LASAA repository:
\ttt{juliet\_prompt.txt},
\ttt{formai\_prompt.txt},
\ttt{svcomp\_prompt.txt}.

\paragraph{Metrics.}
Viewing adjudication as a binary classifier whose positive class is ``a real
flaw is present'' (Table~\ref{tab:confusion}), we report two
class-conditional rates:
\begin{itemize}
\item \emph{Recall} (sensitivity) is the fraction of real-bug alerts that the tool returns for human attention.
\item \emph{Specificity} is the fraction of false alarms that the tool filters out.
\end{itemize}
For both rates, we treat \ttt{dependent} and
\ttt{uncertain} verdicts as positive: such an alert is returned
for human attention rather than discarded, so it is grouped with \ttt{true},
whereas only a \ttt{false} verdict discards the alert.

We report recall together with specificity because, unlike precision, neither
depends on the actually-true/dep \itc{prevalence}
(the percentage of alerts that are actually true/dep), which in
our benchmarks is an artifact of benchmark construction rather than an
operational rate.
We will use ``$\pi$'' to denote the actually-true/dep prevalence.
\itc{Precision} (positive predictive value) answers the question
``Of all alerts that the LLM-based tool returns for human attention, what percent are actually true/dep?''.
Computed at an operational alert prevalence, precision is of great
relevance to analysts/developers dealing with alerts, because it gives the
probability that a given returned alert indicates a real bug.
Because precision is a quantity the analyst ultimately cares about, we also
report it, but at two prevalences to make its prevalence-dependence explicit:
$\pi$ equal to the benchmark's own prevalence and $\pi{=}10\%$, a realistic%
\footnote{
  E.g., Du et al.~\cite{du2026tencent} note that in their Tencent study,
  ``the false positive rate of static warnings is higher than 90\%''
  when not excluding alerts due to incomplete or inaccessible code contexts.
}
operational rate.  We compute precision from recall, specificity, and
prevalence~$\pi$ via
\[
\text{Precision} =
\frac{\text{Recall}\cdot\pi}
     {\text{Recall}\cdot\pi + (1-\text{Specificity})\,(1-\pi)},
\]
treating, as above, \emph{dependent} and \emph{uncertain} verdicts as positive.

In the results tables, each ``Actually $X$'' group of columns collects the alerts whose
ground-truth label (from the benchmark's answer key) is $X$.
(For Juliet and FormAI, the initial answer key is used for the
initial-experiment data tables, and the revised answer key is used for the
revised-experiment data tables; the SV-COMP answer key had no revisions.)
Within such a group, every entry is a
percentage whose denominator is the total number of alerts carrying that ground-truth
label.
The terms ``Wrong'' and ``DepMix'' have the following meanings:
\begin{itemize}
\item \emph{Wrong} under ``actually true/dep'': a true or dependent alert discarded as false.
\item \emph{Wrong} under ``actually false'': a false alert adjudicated as true or dependent.
\item \emph{DepMix}: The LLM adjudicated an actually-true alert as dependent or vice versa.
\end{itemize}
The \emph{dependent} option was
offered only on FormAI; Juliet and SV-COMP use a binary answer key, so their
\emph{DepMix} is identically zero and is omitted.

\begin{table}[t]
\centering\small
\caption{Adjudication viewed as a binary classifier.  Here, we lump dependent alerts in with true alerts.}
\label{tab:confusion}
\begin{tabular}{l|cc}
\toprule
Ground truth & LLM says ``true'' & LLM  says ``false'' \\
\midrule
True (real bug) & TP & FN \\
False           & FP & TN \\
\bottomrule
\end{tabular}
\end{table}

\subsection{Benchmarks}
We evaluate on three suites with differing characteristics.
\emph{Juliet}~\cite{juliet} is a large synthetic C/C++ suite (over 60{,}000
test cases across 118+ CWEs) built to exercise static analyzers; each test case
contains a ``good'' and a ``bad'' variant. \emph{FormAI}~(v2)~\cite{tihanyi2023formai,tihanyi2024secure}
contains 331{,}000 compilable C programs generated by various LLMs, with
vulnerabilities identified by the ESBMC model
checker~\cite{cordeiro2012esbmc,gadelha2018esbmc,menezes2024esbmc}.
\emph{SV-COMP}~\cite{svcomp2024}
provides small, mostly hand-crafted programs annotated with formal properties;
we use a subset with known answers for the memory-safety property
\texttt{valid-deref} (every pointer dereference is valid).

For Juliet and SV-COMP, we ran 10 trials of the \ttt{orig} query,%
\footnote{Recall from Section~\ref{sec:MMM}: The term ``\ttt{orig} query'' is
used in contrast to the term ``LRE query''; it \itc{does~not} refer exclusively
to the queries in the initial experiments in contrast to the revised experiments.
Both the initial experiments and the revised experiments have \ttt{orig} queries.}
except on FormAI we ran the large frontier models \ttt{gpt-5.4}, \ttt{gpt-5.5}, and \ttt{opus-4.8}
with only 5 trials each.
In the data tables, the rows where both CC and LRE are disabled show the
average of these trials.
A single replication of CC-only and CC+LRE was performed for Juliet and SV-COMP;
the CC-only and CC+LRE rows show the results of those replications.

For FormAI, we performed 10 replications of CC-only and CC+LRE for
\ttt{o4-mini}, \ttt{gpt-oss-120b}, and \ttt{gpt-oss-20b}.
For those models, the CC-only and CC+LRE rows show the average of those 10
replications (each involving 10 independent trials of the \ttt{orig} query),
and the rows where neither CC nor LRE are enabled show the average of the
$10 \times 10 = 100$ trials of the \ttt{orig} query.

To quantify uncertainty due to the LLM's nondeterminism on the particular
subsets we evaluated, we also computed 99\% pooled exact
(Clopper--Pearson) confidence intervals for recall and specificity from the
success and trial counts underlying the revised-result tables.  These intervals
are for repeated runs on the sampled alerts, not for generalizing from the
sampled alerts to the rest of a benchmark.  Pooling treats the trial-level success
probability as homogeneous; when success probabilities differ across fixed test
cases, this binomial variance is conservative, since
$\sum_i p_i(1-p_i) \le n\,\bar p(1-\bar p)$.

\subsection{Ground truth and its difficulty}
\label{sec:groundtruth}
Scoring an adjudicator requires knowing the right answer for each alert, and
obtaining that ground truth proved to be one of the harder parts of this work.
FormAI's supplied answer key was produced automatically using the ESBMC model
checker and contains many false alarms incorrectly labeled as true, due to
imprecision in modeling system library functions. We therefore
manually adjudicated a sample of FormAI alerts ourselves, using the three-way
true/false/dependent distinction, and scored the LLMs against those
unpublished adjudications.

Even careful manual adjudications were not final. Modern LLMs are now good enough
that, when an LLM disagrees with a manual adjudication, it is sometimes the manual adjudication
that is wrong: the LLM's written justification (and for true positives, the
dynamic trigger test of Sec.~\ref{sec:trigger}) repeatedly revealed genuine
mistakes in our own adjudications and convinced us to revise them.
Table~\ref{tab:revised-verdicts-formai} lists the FormAI alerts whose manual
adjudication we changed after reviewing LLM output or trigger results.

\subsection{Juliet}
\paragraph{Sanitization.}
Identifiers and comments in raw Juliet files often reveal the answer, so
after splitting each case into its \ttt{GOOD} and \ttt{BAD} variants we sanitize them:
\begin{itemize}
\item
    We remove comments.
\item
    We rename functions/classes defined in the sample,
    variables whose names begin with ``CWE'' or end with ``Global'',
    user-defined types ending in ``Type'', and namespaces beginning with ``CWE''.
\item
    We replace occurrences of ``good'' and ``bad'' (in all lowercase, in all
    uppercase, and in title case) in identifiers and in string literals.
\end{itemize}

It is not obvious how to automatically identify the line number to flag in all
Juliet test cases, so our prompt asked the LLM to determine whether the
specified CWE is present anywhere in the test-case half.
(Juliet metadata identifies a line number for \ttt{BAD} test cases but not for
\ttt{GOOD} test cases.)

\paragraph{Initial results.}
Table~\ref{tab:juliet-v1} reports class-conditional results on a random sample of
100 test cases, drawn uniformly without replacement from the full suite (no
stratification by CWE, so each CWE appears in proportion to its prevalence).
Each case contributes a true (``bad'') half and a false (``good'') half, giving
100 actually-true and 100 actually-false alerts; the sample spans 40 distinct
CWEs.\footnote{Per-CWE composition (CWE: count): CWE-15:1, CWE-23:4,
CWE-36:3, CWE-78:14, CWE-121:7, CWE-122:7, CWE-124:1, CWE-126:2, CWE-127:2,
CWE-134:2, CWE-190:4, CWE-191:5, CWE-195:2, CWE-197:1, CWE-252:1, CWE-253:2,
CWE-284:1, CWE-325:1, CWE-369:1, CWE-390:1, CWE-400:1, CWE-401:3, CWE-404:1,
CWE-415:2, CWE-416:1, CWE-427:3, CWE-457:2, CWE-476:1, CWE-590:2, CWE-605:1,
CWE-606:1, CWE-617:3, CWE-665:2, CWE-675:1, CWE-680:1, CWE-758:1, CWE-761:1,
CWE-762:7, CWE-775:1, CWE-789:3.}


\begin{table*}[t]
\centering\small
\caption{Results from initial Juliet experiments: class-conditional results on
a random sample of 100 cases (no consistency check; no \emph{dependent}
option). Each case contributes a true (``bad'') and a false (``good'') half.
The \emph{Precision} is provided at the benchmark prevalence ($\pi{=}50\%$)
and at $\pi{=}10\%$.}
\label{tab:juliet-v1}
\renewcommand{\arraystretch}{1.1}
\begin{tabular}{l p{0.5em} ccc p{0.5em} ccc p{0em} cc}
\toprule
& & \multicolumn{3}{c}{Actually true (100)} & & \multicolumn{3}{c}{Actually false (100)} & & \multicolumn{2}{c}{Precision} \\
\cmidrule(lr){3-5}\cmidrule(lr){7-9}\cmidrule(lr){11-12}
LLM          & & Recall & Wrong & Uncert & & Spec.  & Wrong  & Uncert & & $\pi{=}50\%$ & $\pi{=}10\%$ \\
\midrule
o4-mini      & & 97.7\% & 2.3\% & 0.0\%  & & 98.2\% & 1.8\%  & 0.0\%  & & 98.2\%       & 85.8\% \\
gpt-oss-120b & & 96.8\% & 3.2\% & 0.0\%  & & 97.4\% & 2.6\%  & 0.0\%  & & 97.4\%       & 80.5\% \\
gpt-oss-20b  & & 96.1\% & 3.9\% & 0.0\%  & & 96.3\% & 3.7\%  & 0.0\%  & & 96.3\%       & 74.3\% \\
gpt-4o       & & 98.0\% & 2.0\% & 0.0\%  & & 90.0\% & 10.0\% & 0.0\%  & & 90.7\%       & 52.1\% \\
\bottomrule
\end{tabular}
\end{table*}

\begin{table*}[t]
\centering\small
\caption{Results from revised Juliet experiments ($N{=}10$ trials per phase),
using the refined prompts and one corrected answer-key entry (a test case originally
labeled as CWE-122 that we relabeled as CWE-121). ``CC'' is the consistency-check threshold (``no'' = no consistency check);
The \emph{Precision} is provided at the benchmark prevalence ($\pi{=}50\%$)
and at $\pi{=}10\%$.}
\label{tab:juliet-v2}
\renewcommand{\arraystretch}{1.1}

\begin{tabular}{l c c p{0.5em} ccc p{0.5em} ccc p{0em} cc}
\toprule
& & & & \multicolumn{3}{c}{Actually true (100)} & & \multicolumn{3}{c}{Actually false (100)} & & \multicolumn{2}{c}{Precision} \\
\cmidrule(lr){5-7}\cmidrule(lr){9-11}\cmidrule(lr){13-14}

LLM          & LRE & CC   & & Recall & Wrong & Uncert & & Spec.  & Wrong & Uncert & & $\pi{=}50\%$ & $\pi{=}10\%$ \\
\midrule
o4-mini      & no  & no   & & 98.1\% & 1.9\% & 0.0\%  & & 98.8\% & 1.2\% & 0.0\%  & & 98.8\%       & 90.1\% \\
o4-mini      & no  & 80\% & & 99.0\% & 1.0\% & 2.0\%  & & 99.0\% & 1.0\% & 0.0\%  & & 99.0\%       & 91.7\% \\
o4-mini      & yes & 80\% & & 97.0\% & 3.0\% & 0.0\%  & & 99.0\% & 1.0\% & 0.0\%  & & 99.0\%       & 91.5\% \\
\addlinespace
gpt-oss-120b & no  & no   & & 97.8\% & 2.2\% & 0.0\%  & & 98.3\% & 1.7\% & 0.0\%  & & 98.3\%       & 86.5\% \\
gpt-oss-120b & no  & 80\% & & 98.0\% & 2.0\% & 0.0\%  & & 97.0\% & 1.0\% & 2.0\%  & & 97.0\%       & 78.4\% \\
gpt-oss-120b & yes & 80\% & & 97.0\% & 3.0\% & 0.0\%  & & 96.0\% & 3.0\% & 1.0\%  & & 96.0\%       & 72.9\% \\
\addlinespace
gpt-oss-20b  & no  & no   & & 97.6\% & 2.4\% & 0.0\%  & & 96.7\% & 3.3\% & 0.0\%  & & 96.7\%       & 76.7\% \\
gpt-oss-20b  & no  & 80\% & & 99.0\% & 1.0\% & 1.0\%  & & 96.0\% & 1.0\% & 3.0\%  & & 96.1\%       & 73.3\% \\
gpt-oss-20b  & yes & 80\% & & 96.0\% & 4.0\% & 0.0\%  & & 96.0\% & 2.0\% & 2.0\%  & & 96.0\%       & 72.7\% \\
\addlinespace
gpt-4o       & no  & no   & & 97.7\% & 2.3\% & 0.0\%  & & 91.9\% & 8.1\% & 0.0\%  & & 92.3\%       & 57.3\% \\
gpt-4o       & no  & 80\% & & 99.0\% & 1.0\% & 2.0\%  & & 86.0\% & 1.0\% & 13.0\% & & 87.6\%       & 44.0\% \\
gpt-4o       & yes & 80\% & & 99.0\% & 1.0\% & 0.0\%  & & 89.0\% & 6.0\% & 5.0\%  & & 90.0\%       & 50.0\% \\
\bottomrule
\end{tabular}
\end{table*}

\paragraph{Revised results.} 
For the revised Juliet experiment round, we made two changes:
\begin{itemize}
\item
    For CWE-325 (``Missing Required Cryptographic Step'') alerts, we added the following clarification to the prompt: ``This is NOT about CWE-329 (Predictable IV for CBC) or CWE-1204 (Weak IV)''.
\item
    For \verb`CWE122..._c_src_wchar_t_cpy_13` (\ttt{BAD} version): we changed the CWE category from CWE-122 (``Heap Based Buffer Overflow'') to CWE-121 (``Stack Based Buffer Overflow''), because the overflowed buffer is actually on the stack, not on the heap.  (This was a mistake in Juliet.)
    We noticed this because o4-mini consistently contradicted the answer key
    on this test case; after reading the LLM's explanation and re-adjudicating
    the case manually with that explanation in mind, we concluded that the LLM
    was right and the answer key was wrong.
\end{itemize}

Table~\ref{tab:juliet-v2} reports the results.
The reasoning LLMs adjudicate both halves well, whereas the
larger but non-reasoning \texttt{gpt-4o} preserves real flaws about as well as
they do (97--98\% recall) yet lets through far more false positives
($\approx$90\% specificity vs.\ 96--99\%). That is, \texttt{gpt-4o}'s lower
accuracy is almost entirely false alarms---the comparatively safe error.

Interestingly, enabling LRE tends to \itc{hurt} performance on Juliet, unlike
FormAI and SV-COMP, where it tends to help.  From a preliminary investigation,
it seems that there are a few points on which the original prompt is unclear,
and LRE latches onto the wrong interpretation.  As future work, we will
investigate whether further tweaks to the prompt can improve LRE performance.

\begin{table*}[t]
\centering
\caption{Changes in manual adjudications for FormAI alerts.}
\label{tab:revised-verdicts-formai}
\renewcommand{\arraystretch}{1.2}
\begin{tabularx}{\textwidth}{@{}lcc>{\raggedright\arraybackslash}X@{}}
\toprule
& \multicolumn{2}{c}{Verdict} & \\
Alert  & Old   & New   & Notes \\
\midrule
068671 & True  & Dep   & If no overflow on line~54, then $\texttt{julianDay} \le \texttt{INT\_MAX} - 32083$, so line~56 can't overflow. \\
364317 & Dep   & True  & Trigger: Negative value for \texttt{size}. \\
576500 & Dep   & True  & Trigger: \texttt{n=0} and \texttt{malloc} fails and \texttt{range > 0}. \\
586756 & False & True  & Trigger: \texttt{p=-32767}, \texttt{q=65537}, \texttt{e=1}. \\
700724 & False & Dep   & Trigger: \texttt{f\_blocks =} $2^{63}$, \texttt{f\_bfree = 1}, \texttt{f\_frsize = 1} for 64-bit \texttt{long long}. \\
659854 & Dep   & False & \texttt{scanf} is given an invalid pointer, so wrong alert category. \\
867712 & True  & False & \texttt{scanf} is given an invalid pointer, so wrong alert category. \\
\bottomrule
\end{tabularx}
\end{table*}

\paragraph{Memorization risk.}
Juliet was certainly in these LLMs' training data, so high accuracy
could in principle reflect memorization rather than genuine reasoning. This
is why our headline robustness claims rest on FormAI (see below).

\emph{Our original reasons for discounting memorization.}
We originally had three reasons to believe that the LLMs were not simply
overfitting to Juliet:
(1)~We sanitize each case before querying (Sec.~\ref{sec:eval}, ``Sanitization''),
removing comments and the identifiers, type names, and namespaces that
otherwise reveal the answer; we reasoned that this stripped the surface cues
an LLM would need to recall the case.
(2)~The small reasoning LLM \ttt{gpt-oss-20b} did noticeably better than the
larger non-reasoning \texttt{gpt-4o} (96.7\% vs.\ 91.9\% specificity)%
\footnote{With 967 vs.\ 919 out of 1000 pooled trials
(100 test cases, each repeated 10 times),
we get a p-value of $p \approx 4.3\times10^{-6}$ using Fisher's exact test,
two-sided. The $p$-value here tests only whether this result (gpt-oss-20b has
better specificity than gpt-4o on our subset of Juliet) is stable against the
LLMs' sampling randomness, i.e., whether we ran enough trials that the same
result would hold in the large-trial limit on these same test cases.  We
are~not testing whether it generalizes to the rest of Juliet.}.
Pure memorization and overfitting would not predict that ordering.
(3)~One of the LLMs' ``wrong'' answers is actually the LLM \emph{correcting}
the Juliet answer key: the relabeling of \verb`CWE122..._c_src_wchar_t_cpy_13`
from CWE-122 to CWE-121 described under ``Revised results'' above.
This provided some evidence the LLM was reasoning about
the code rather than reciting an adjudication.

\emph{Why these reasons are weaker than we thought.}
We directly tested Reason~(1), and it does not hold up: as shown next, o4-mini
(and to a lesser extent gpt-oss-120b) can often recover the original test-case
identity even after sanitization.
Reason~(2) still provides some evidence that capability (not overfitting) is
involved, but it is relatively weak evidence.
A counterargument to Reason~(3) is that there is public discussion available on
the Internet (and presumably in the LLM's training data) of specific errors in
the Juliet answer key, so an LLM may simply have memorized the correction
rather than derived it.
We learned of this discussion only after making the correction ourselves:
some time after we had relabeled \verb`CWE122..._cpy_13`, it was suggested to us
that corrections to Juliet might already be available online.
A search (conducted with GPT-5.5-high) revealed that Sti{\'e}venart et
al.~\cite{stievenart2021-juliet-cwe122} report the same kind of
mislabeling (a test case where the overflowed buffer is actually on the stack
but is mislabeled as ``CWE-122: Heap Based Buffer Overflow'')
in a very similar test case,
\verb`CWE122..._c_src_wchar_t_cat_03`.
So although we reached our correction independently of the published one, an
LLM whose training data discusses the mislabeling of \verb`CWE122..._cat_03` could
plausibly transfer the correction to \verb`CWE122..._cpy_13` without reasoning
about the code.

\emph{Measuring recognition directly.} To estimate how much of the case
identity survives sanitization, we ran a filename-reconstruction probe. For
each of the 200 test-case halves in our 100-case sample (each case has a good
and a bad half), we gave the LLM only the sanitized code plus the CWE number
(the same CWE the adjudication prompt receives) and asked it to reconstruct
the original Juliet filename, which encodes a \emph{functional variant} (e.g.,
\texttt{char\_alloca\_memcpy}) and a numeric \emph{flow variant}. We ran the
probe with o4-mini, gpt-oss-120b, and gpt-oss-20b, then used a separate LLM
(GPT-5.5-high) to score each guess against the true filename: a
\emph{functional} score of 0--3 (3~=~exact up to punctuation/case; 2~=~more
than half the parts correct; 1~=~at least one part correct; 0~=~unrelated) and
a binary \emph{flow} score (1 if the flow-variant number matches, 0 otherwise).
One author double-checked by manually scoring a simple random sample of 20
test-case halves.  On 19 of 20, the manual score matched GPT-5.5's score.
The one difference was for a test-case half where the correct answer for the
functional variant was \qt{\ttt{w32\_char\_CreateWindowStation}} but gpt-oss-120b
answered \qt{\ttt{w32CreateWindowStation}}; the manual score was 1 but
GPT-5.5 gave a score of 2.
Note that a functional-variant score of 1 is somewhat easy for the LLM to
achieve without memorizing a lot of things specific to individual test cases;
e.g., in all 27 test cases where a file contains ``\verb+wchar_t+'', the
functional-variant portion of the filename also contains ``\verb+wchar_t+''.

\begin{table}[t]
\centering\small
\caption{Filename-reconstruction probe on the sanitized 100-case Juliet
sample (200 test-case halves). ``Func=3'' = exact functional variant;
``Func~$\ge$2'' = exact or largely correct; ``Func~$\ge$1'' = at least one
part correct; ``Flow'' = exact flow-variant number. Higher values mean more
case identity survives sanitization, i.e., greater memorization risk.}
\label{tab:juliet-memorization}
\begin{tabular}{lrrrr}
\toprule
LLM & Func=3 & Func$\ge$2 & Func$\ge$1 & Flow \\
\midrule
o4-mini        & 68\% & 91\% & 97\% & 48\% \\
gpt-oss-120b   & 45\% & 84\% & 96\% & 16\% \\
gpt-oss-20b    & 10\% & 47\% & 94\% &  6\% \\
\bottomrule
\end{tabular}
\end{table}

Table~\ref{tab:juliet-memorization} reports the results, and they are
sobering. The LLM o4-mini reconstructs the exact functional variant for 68\% of the
halves and gets it largely or exactly right for 91\%; even the flow-variant
number is exactly recovered 48\% of the time.
The LLM gpt-oss-120b is fairly close behind on the functional variant
(84\% at score~$\ge$2). Only gpt-oss-20b shows weak recognition (47\% at
score~$\ge$2, 10\% exact). We therefore judge the memorization risk as
\emph{high} for o4-mini, \emph{medium-high} for gpt-oss-120b, and \emph{low}
for gpt-oss-20b.

\emph{Flipped-verdict variants.} High recognition of a case's
\emph{identity} (Table~\ref{tab:juliet-memorization}) does not by itself show
that the LLM is reciting a memorized \emph{verdict}: a model could recognize
the template yet still be reasoning about the code in front of it. To separate
these two possibilities, we created \emph{flipped-verdict} variants of 20 Juliet
test cases (randomly selected from our subset of 100 test cases),
spanning 13 distinct CWEs. In each variant we introduced a flaw of the specified CWE type
into the \ttt{GOOD} half and repaired the flaw in the \ttt{BAD} half, so that
the correct verdict of each half is the opposite of the original.\footnote{The
guards (\ttt{\#ifndef OMITGOOD} / \ttt{\#ifndef OMITBAD}) and the splitting and
sanitization pipeline are unchanged; only the code of each half was
edited.} A model that has merely memorized which template-half carries the flaw
would systematically get these variants \emph{wrong} (calling the now-flawed
\ttt{GOOD} half false and the now-repaired \ttt{BAD} half true), whereas a model
that actually analyzes the code should still adjudicate them correctly.

One author wrote 3 of the variants as examples and a large frontier LLM
(Claude Opus~4.8, via Claude Code) wrote the remaining 17, all of which we then reviewed
manually.
To try to keep the variants from being solvable by pattern-matching to a remembered
template, we instructed Claude that the edited \ttt{GOOD} half should not
closely resemble a \ttt{BAD} half of any existing Juliet test case, and vice versa.
We did not fully achieve the dissimilarity goal in every case (for
some CWEs, there are few natural alternatives), but the resulting variants are on
the whole distinct from the stock Juliet templates.
We also ran the filename-reconstruction probe on the transformed test-case
halves, for two purposes: to check that our edits had not simply turned a
case into a near-copy of a \emph{different} test case already present in
Juliet, and to check whether the variants were still recognizable at all.
The LLM \ttt{o4-mini} reconstructed the functional-variant part of the
\emph{original} filename for 29 of the 40 transformed halves.  In other
words, our edits do not defeat recognition: o4-mini can still usually tell
which Juliet template a variant came from.

We next ran o4-mini (the model with the highest measured memorization risk) on all
40 test-case halves, with 5 trials per half and the consistency check. It
adjudicated every half correctly (a perfect score: all 20 now-flawed
\ttt{GOOD} halves judged true and all 20 now-repaired \ttt{BAD} halves judged
false). This is direct evidence that o4-mini's Juliet accuracy reflects analysis
of the code rather than recall of a memorized verdict.

Although the flipped-verdict experiment provides some evidence that the LLMs
aren't simply overfitting to Juliet, it doesn't definitively resolve the question.
Our FormAI experiments provide better evidence that the LLM's performance isn't due
to overfitting, because the adjudications we score against are our own
unpublished manual adjudications.  
Additionally, the three-way true/false/dependent
distinction that we use in the FormAI experiments
is less popular (and therefore less represented in the LLM's training data)
than a binary true/false scheme and therefore provides additional evidence
of general LLM reasoning capabilities for code.

\subsection{FormAI}
\label{sec:eval:formai}
\paragraph{Initial results.}
We randomly sampled 100 items in the FormAI database.  Of these, 15 were
labeled ``\ttt{UNKNOWN (time out)}'' and 3 were labeled ``\ttt{NON-VULNERABLE}'' (meaning
no vulnerability of any type was found on any line).
We discarded these, leaving only alerts labeled \ttt{VULNERABLE}.
We then discarded the 3 alerts whose error type matched
``\ttt{arithmetic overflow on floating-point ieee\_(mul|div)}''.
This left 79 FormAI vulnerability reports, which we manually adjudicated
to establish ground truth (Sec.~\ref{sec:groundtruth}).
Our initial manual adjudications were: 31 actually true, 20 actually dependent, and 28 actually false.
We ran 10 trials per case.
Table~\ref{tab:formai-v1} reports results both with and without a
consistency check, using a
per-LLM threshold chosen as the smallest value under which less than 5\% of the
79 adjudications were wrong (here, counting mixups between \ttt{true} and
\ttt{dependent} as wrong).

In contrast to Juliet and SV-COMP, we do have flagged line numbers for all
FormAI test cases (both actually-true alerts and actually-false alerts)%
\footnote{As discussed above, our subset of FormAI consists only of alerts
that were labeled as true positives by the FormAI metadata, which is why we
have line numbers.  Many of the FormAI labels are wrong, which is why we have a
decent number of actually-false test cases after doing our own manual
adjudication.},
so our
prompt asked the LLM to determine whether the specified CWE is
present on the specified line.

\begingroup
\setlength{\tabcolsep}{3pt} 

\begin{table*}[t]
\centering\small
\caption{Results from initial FormAI experiments.
\emph{Precision} is provided at the benchmark prevalence
($\pi{=}65\%$) and at $\pi{=}10\%$.}
\label{tab:formai-v1}
\renewcommand{\arraystretch}{1.1}
\begin{tabular}{l cc p{\dimexpr1.5em-2\tabcolsep\relax} cccc p{\dimexpr1.5em-2\tabcolsep\relax} ccc p{\dimexpr1.5em-2\tabcolsep\relax} cc}
\toprule
& & & & \multicolumn{4}{c}{Actually true/dep (51)} & & \multicolumn{3}{c}{Actually false (28)} & & \multicolumn{2}{c}{Precision} \\
\cmidrule(lr){5-8}\cmidrule(lr){10-12}\cmidrule(lr){14-15}
LLM          & LRE & CC   & & Recall  & Wrong & DepMix & Uncert & & Spec.  & Wrong  & Uncert & & $\pi{=}65\%$ & $\pi{=}10\%$ \\
\midrule
o4-mini      & no  & no   & & 95.1\%  & 4.9\% & 6.7\%  & 0.0\%  & & 93.2\% & 6.8\%  & 0.0\%  & & 96.2\%       & 60.8\% \\
o4-mini      & no  & 80\% & & 96.1\%  & 3.9\% & 2.0\%  & 9.8\%  & & 89.3\% & 0.0\%  & 10.7\% & & 94.2\%       & 49.9\% \\
\addlinespace
gpt-oss-120b & no  & no   & & 95.3\%  & 4.7\% & 9.4\%  & 0.0\%  & & 87.1\% & 12.9\% & 0.0\%  & & 93.1\%       & 45.1\% \\
gpt-oss-120b & no  & 90\% & & 100.0\% & 0.0\% & 2.0\%  & 23.5\% & & 78.6\% & 0.0\%  & 21.4\% & & 89.5\%       & 34.2\% \\
\addlinespace
gpt-oss-20b  & no  & no   & & 92.9\%  & 7.1\% & 16.5\% & 0.0\%  & & 83.9\% & 16.1\% & 0.0\%  & & 91.3\%       & 39.1\% \\
gpt-oss-20b  & no  & 90\% & & 96.1\%  & 3.9\% & 0.0\%  & 43.1\% & & 75.0\% & 3.6\%  & 21.4\% & & 87.5\%       & 29.9\% \\
gpt-oss-20b  & yes & 90\% & & 96.1\%  & 3.9\% & 0.0\%  & 13.7\% & & 78.6\% & 7.1\%  & 14.3\% & & 89.1\%       & 33.3\% \\
\bottomrule
\end{tabular}
\end{table*}

\begin{table*}[tp]
\centering\small
\caption{Results from revised FormAI experiments, using the refined prompts and
our revised manual adjudications.
In the ``CC'' column, \qt{\ttt{maj}} denotes the plain majority-vote baseline
of Sec.~\ref{sec:MMM} (applied to the phase-1 trials, with no LRE).
The \emph{Precision} is provided at the benchmark
prevalence ($\pi{=}64\%$) and at $\pi{=}10\%$.}
\label{tab:formai-v2}
\renewcommand{\arraystretch}{1.1}
\begin{tabular}{l cc p{\dimexpr1.5em-2\tabcolsep\relax} cccc p{\dimexpr1.5em-2\tabcolsep\relax} ccc p{\dimexpr1.5em-2\tabcolsep\relax} cc}
\toprule
& & & & \multicolumn{4}{c}{Actually true/dep (49)} & & \multicolumn{3}{c}{Actually false (27)} & & \multicolumn{2}{c}{Precision} \\
\cmidrule(lr){5-8}\cmidrule(lr){10-12}\cmidrule(lr){14-15}
LLM          & LRE & CC   & & Recall  & Wrong & DepMix & Uncert & & Spec.   & Wrong  & Uncert & & $\pi{=}64\%$ & $\pi{=}10\%$ \\
\midrule
o4-mini      & no  & no   & & 96.6\%  & 3.4\% & 9.2\%  & 0.0\%  & & 93.5\%  & 6.5\%  & 0.0\%  & & 96.4\%       & 62.4\% \\
o4-mini      & no  & maj  & & 96.9\%  & 3.1\% & 10.0\% & 0.0\%  & & 93.0\%  & 7.0\%  & 0.0\%  & & 96.1\%       & 60.5\% \\
o4-mini      & no  & 70\% & & 98.4\%  & 1.6\% & 5.1\%  & 9.2\%  & & 90.4\%  & 1.9\%  & 7.8\%  & & 94.8\%       & 53.2\% \\
o4-mini      & no  & 80\% & & 99.2\%  & 0.8\% & 3.9\%  & 12.0\% & & 88.9\%  & 0.4\%  & 10.7\% & & 94.1\%       & 49.8\% \\
o4-mini      & yes & 80\% & & 99.4\%  & 0.6\% & 6.1\%  & 2.9\%  & & 94.8\%  & 1.1\%  & 4.1\%  & & 97.1\%       & 68.0\% \\
o4-mini      & yes & 70\% & & 99.4\%  & 0.6\% & 6.5\%  & 1.8\%  & & 96.7\%  & 1.1\%  & 2.2\%  & & 98.1\%       & 76.8\% \\
\addlinespace
gpt-oss-120b & no  & no   & & 98.3\%  & 1.7\% & 11.4\% & 0.0\%  & & 94.4\%  & 5.6\%  & 0.0\%  & & 96.9\%       & 66.1\% \\
gpt-oss-120b & no  & maj  & & 99.4\%  & 0.6\% & 12.0\% & 0.0\%  & & 97.4\%  & 2.6\%  & 0.0\%  & & 98.6\%       & 81.0\% \\
gpt-oss-120b & no  & 70\% & & 99.8\%  & 0.2\% & 6.3\%  & 10.6\% & & 94.4\%  & 0.4\%  & 5.2\%  & & 97.0\%       & 66.6\% \\
gpt-oss-120b & no  & 80\% & & 100.0\% & 0.0\% & 4.7\%  & 14.5\% & & 91.1\%  & 0.0\%  & 8.9\%  & & 95.2\%       & 55.6\% \\
gpt-oss-120b & yes & 80\% & & 100.0\% & 0.0\% & 7.8\%  & 2.0\%  & & 96.3\%  & 0.7\%  & 3.0\%  & & 98.0\%       & 75.0\% \\
gpt-oss-120b & yes & 70\% & & 100.0\% & 0.0\% & 8.0\%  & 1.0\%  & & 97.8\%  & 0.7\%  & 1.5\%  & & 98.8\%       & 83.3\% \\
\addlinespace
gpt-oss-20b  & no  & no   & & 96.9\%  & 3.1\% & 15.3\% & 0.0\%  & & 93.6\%  & 6.4\%  & 0.0\%  & & 96.4\%       & 62.6\% \\
gpt-oss-20b  & no  & maj  & & 98.4\%  & 1.6\% & 8.4\%  & 0.0\%  & & 93.0\%  & 7.0\%  & 0.0\%  & & 96.1\%       & 60.8\% \\
gpt-oss-20b  & no  & 70\% & & 98.8\%  & 1.2\% & 3.1\%  & 16.7\% & & 92.6\%  & 2.2\%  & 5.2\%  & & 96.0\%       & 59.7\% \\
gpt-oss-20b  & no  & 80\% & & 99.6\%  & 0.4\% & 1.8\%  & 29.8\% & & 92.2\%  & 0.7\%  & 7.0\%  & & 95.8\%       & 58.7\% \\
gpt-oss-20b  & yes & 80\% & & 99.4\%  & 0.6\% & 4.9\%  & 7.1\%  & & 97.0\%  & 1.1\%  & 1.9\%  & & 98.4\%       & 78.8\% \\
gpt-oss-20b  & yes & 70\% & & 99.0\%  & 1.0\% & 5.3\%  & 4.5\%  & & 97.4\%  & 1.5\%  & 1.1\%  & & 98.5\%       & 80.9\% \\
\addlinespace
gpt-4o       & no  & no   & & 95.1\%  & 4.9\% & 27.6\% & 0.0\%  & & 90.0\%  & 10.0\% & 0.0\%  & & 94.4\%       & 51.4\% \\
gpt-4o       & no  & maj  & & 95.9\%  & 4.1\% & 34.7\% & 0.0\%  & & 96.3\%  & 3.7\%  & 0.0\%  & & 97.9\%       & 74.2\% \\
gpt-4o       & no  & 80\% & & 95.9\%  & 4.1\% & 10.2\% & 36.7\% & & 77.8\%  & 3.7\%  & 18.5\% & & 88.5\%       & 32.4\% \\
gpt-4o       & yes & 80\% & & 98.0\%  & 2.0\% & 18.4\% & 12.2\% & & 88.9\%  & 3.7\%  & 7.4\%  & & 94.0\%       & 49.5\% \\
\addlinespace
gpt-5.4      & no  & no   & & 98.8\%  & 1.2\% & 2.4\%  & 0.0\%  & & 99.3\%  & 0.7\%  & 0.0\%  & & 99.6\%       & 93.7\% \\
gpt-5.4      & no  & 80\% & & 100.0\% & 0.0\% & 2.0\%  & 2.0\%  & & 100.0\% & 0.0\%  & 0.0\%  & & 100.0\%      & 100.0\% \\
gpt-5.4      & yes & 80\% & & 98.0\%  & 2.0\% & 2.0\%  & 2.0\%  & & 100.0\% & 0.0\%  & 0.0\%  & & 100.0\%      & 100.0\% \\
\addlinespace
gpt-5.5      & no  & no   & & 100.0\% & 0.0\% & 1.2\%  & 0.0\%  & & 100.0\% & 0.0\%  & 0.0\%  & & 100.0\%      & 100.0\% \\
gpt-5.5      & no  & 80\% & & 100.0\% & 0.0\% & 0.0\%  & 2.0\%  & & 100.0\% & 0.0\%  & 0.0\%  & & 100.0\%      & 100.0\% \\
gpt-5.5      & yes & 80\% & & 100.0\% & 0.0\% & 0.0\%  & 0.0\%  & & 100.0\% & 0.0\%  & 0.0\%  & & 100.0\%      & 100.0\% \\
\addlinespace
opus-4.8     & no  & no   & & 98.4\%  & 1.6\% & 0.0\%  & 0.0\%  & & 98.5\%  & 1.5\%  & 0.0\%  & & 99.2\%       & 88.1\% \\
opus-4.8     & no  & 80\% & & 100.0\% & 0.0\% & 0.0\%  & 2.0\%  & & 100.0\% & 0.0\%  & 0.0\%  & & 100.0\%      & 100.0\% \\
opus-4.8     & yes & 80\% & & 100.0\% & 0.0\% & 0.0\%  & 2.0\%  & & 100.0\% & 0.0\%  & 0.0\%  & & 100.0\%      & 100.0\% \\
\bottomrule
\end{tabular}
\end{table*}

\endgroup

\begin{table}[tb]
\centering\small
\caption{Alert count for revised FormAI experiment}
\label{tab:formai-counts}
\renewcommand{\arraystretch}{1.1}
\begin{tabular}{lrrr}
\hline
Alert category            & True & Dep & False\\
\hline
arithmetic overflow        &  5  &   2 &    1 \\
array bounds violated      &  4  &   0 &    0 \\
buffer overflow on scanf   &  6  &   0 &   18 \\
deref: invalidated dyn obj &  0  &   0 &    1 \\
deref: invalid pointer     &  4  &   0 &    5 \\
deref: NULL pointer        & 10  &  17 &    2 \\
division by zero           &  1  &   0 &    0 \\ \hline
TOTAL                      & 30  &  19 &   27 \\
\hline
\end{tabular}
\end{table}

\begin{figure*}
\begin{verbatim}
54    int julianDay = day + (153 * (month + 1)) / 5 + 365 * year + (year / 4) - 32083;

56    int temp = julianDay + 32044;
\end{verbatim}
\caption{Code excerpt from FormAI alert 068671}
\label{fig:julianDay-lines-54-56}
\end{figure*}

\paragraph{Revised results.} 
FormAI's initial prompts were the most ambiguous of the three suites, and they
changed the most when revising the experiments.  We made the following changes
to the prompt:
\begin{enumerate}
\item Clarified that the ``buffer overflow on
\texttt{scanf}'' category applies only when \texttt{scanf} is given a
\emph{valid} pointer (an out-of-bounds memory access caused by an
\emph{invalid} pointer should be indicated by a different alert category).
\item Clarified that an
arithmetic-overflow-on-division alert is not a division-by-zero alert.
(For a given signed-integer bit-width, there is only one case where division
overflows: dividing the most negative representable value by $-1$.)
\item Clarified that undefined behavior in the
\emph{definition} of a value should be flagged at the definition rather than at
a later use, and that tenuously related undefined behavior (UB) should be ignored.
\item Clarified how to treat a signed-integer-overflow alert that can arise only
as a downstream effect of an earlier unsigned-to-signed conversion. When a
large unsigned value is assigned to a signed-integer variable, the high bit is
reinterpreted as a sign bit, so the value becomes negative. A later arithmetic
operation on that now-negative value can then overflow. We specified that such
an overflow should be marked \emph{dependent} on the earlier conversion rather
than reported as a true positive, even though the unsigned-to-signed conversion
is only implementation-defined behavior and not undefined behavior.
\end{enumerate}
In addition to the prompt changes, we removed all three memory-leak alerts
(two initially adjudicated true, one false),
because their text did not adequately identify the allegedly leaked memory.
We also revised
7 manual adjudications that LLM output or trigger results had shown to be
wrong, as shown in Table~\ref{tab:revised-verdicts-formai}; the net effect of
these 7 changes is one more true alert and one fewer dependent alert.
The revised experiment thus had 76 alerts: 30 true, 19 dep, 27 false.
Table~\ref{tab:formai-v2}
reports the revised results: the
consistency check again drives the missed-flaw rate
closer to zero, often at the expense of specificity.

\paragraph{Alert 068671.}
The example of alert 068671 reveals the depth of complexity regarding dependency.
Consider the code that generated this alert, shown in Fig.~\ref{fig:julianDay-lines-54-56}.
The alert specifies a signed integer overflow on line 56.
For the alert to be true, there must also be no signed integer overflow on line 54.
Since addition in C is left-associative,%
\footnote{ISO C23 sub-clause 6.5.1, paragraph 3 and the BNF grammar in sub-clause 6.5.7, paragraph 1.}
line 54 computes \ttt{(julianDay + 32083)} inside the expression,
which overflows, and renders line 56 dependent.

As a simpler example, consider this code:
\begin{samepage}
\begin{verbatim}
      int x = INT_MAX + 1 - 1;
\end{verbatim}
\end{samepage}

If the platform evaluates this as \hbox{\ttt{(INT\_MAX + 1) - 1}}, then overflow occurs,
but if it is evaluated as \hbox{\ttt{INT\_MAX + (1 - 1)}}, then overflow does not occur.
So C integer addition actually violates the associative property of addition!

A fix to prevent overflow
would guarantee that \ttt{julianDay + 32083} does not overflow before executing line 54.
And this will also prevent overflow on line 56.
Therefore the alert should be dependent.
Obviously, the line between dependent vs.\ true alerts is more tricky than we anticipated.

\paragraph{Alert 364317.}
Alert 364317 yields a striking, non-monotonic pattern across model
strength: the weakest model (gpt-4o) and the strongest models (gpt-5.4,
gpt-5.5, and opus-4.8) all consistently give the correct answer (\ttt{true}),
while the middle-tier models give a varying mix of correct and incorrect
answers on the \ttt{orig} query:
\begin{alltt}\small
gpt-oss-20b:   76/100 true, 21/100 dep, 3/100 false
gpt-oss-120b:  37/100 true, 63/100 dep
o4-mini:       31/100 true, 67/100 dep, 2/100 false
\end{alltt}
What is happening here is that there is argument that the alert is dependent that
would be sound if \texttt{size} were always non-negative.
The weakest model (gpt-4o) never raises this argument at all; it merely notes
the missing null-check after \ttt{malloc} and reports \ttt{true}, reaching the
right verdict but with incorrect/incomplete reasoning.
The mid-range models often \emph{do} discover the \ttt{dependent} argument, but many
of their runs stop there and answer \ttt{dependent}, which is why their
verdicts split (o4-mini and gpt-oss-120b mostly \ttt{dependent},
gpt-oss-20b mostly \ttt{true}).
What these runs miss (and what the smartest models consistently discover) is
that the \ttt{dependent} argument collapses when \ttt{size} is negative.
Take \hbox{\ttt{size == -1}}: when \ttt{size} is passed to \ttt{malloc},
it is implicitly cast to a huge unsigned integer,
likely causing \ttt{malloc} to return \ttt{NULL}.
When \ttt{size} is negative, the line flagged by the alert is the first
dereference of the NULL pointer returned by \ttt{malloc} (unlike the case when
\ttt{size} is positive, where there is a prior dereference).

\subsection{SV-COMP}
\paragraph{Initial results.}
We worked with 101 \texttt{valid-deref} samples (42 actually true, 59 actually
false). Table~\ref{tab:svcomp-v1} shows reasonably strong recall across the LLMs, with
o4-mini also reaching high specificity.  As the revised round shows, many of the
incorrect and uncertain adjudications are due to a
question-framing mismatch rather than a reasoning failure.

SV-COMP doesn't specify flaws by line number, so our prompt asked the LLM to
determine whether the specified CWE is present anywhere in the test case.

\begin{table*}[t]
\centering\small
\caption{Results from initial SV-COMP experiments on \texttt{valid-deref} (101
samples: 42 actually true, 59 actually false). ``CC'' is the consistency-check
threshold (``no'' = no consistency check). The suite offers no \emph{dependent} option.
The \emph{Precision} is provided at the benchmark prevalence ($\pi{=}42\%$)
and at $\pi{=}10\%$.}
\label{tab:svcomp-v1}
\renewcommand{\arraystretch}{1.1}
\begin{tabular}{l c p{0.5em} ccc p{0.5em} ccc p{0em} cc}
\toprule
& & & \multicolumn{3}{c}{Actually true (42)} & & \multicolumn{3}{c}{Actually false (59)} & & \multicolumn{2}{c}{Precision} \\
\cmidrule(lr){4-6}\cmidrule(lr){8-10}\cmidrule(lr){12-13}
LLM          & CC   & & Recall & Wrong  & Uncert & & Spec.  & Wrong  & Uncert & & $\pi{=}42\%$ & $\pi{=}10\%$ \\
\midrule
o4-mini      & no   & & 89.0\% & 11.0\% & 0.0\%  & & 99.0\% & 1.0\%  & 0.0\%  & & 98.4\%       & 90.8\% \\
o4-mini      & 90\% & & 90.5\% & 9.5\%  & 2.4\%  & & 98.3\% & 0.0\%  & 1.7\%  & & 97.4\%       & 85.5\% \\
\addlinespace
gpt-oss-120b & no   & & 90.5\% & 9.5\%  & 0.0\%  & & 92.5\% & 7.5\%  & 0.0\%  & & 89.6\%       & 57.3\% \\
gpt-oss-120b & 90\% & & 95.2\% & 4.8\%  & 7.1\%  & & 84.7\% & 0.0\%  & 15.3\% & & 81.6\%       & 40.9\% \\
\addlinespace
gpt-oss-20b  & no   & & 89.0\% & 11.0\% & 0.0\%  & & 86.8\% & 13.2\% & 0.0\%  & & 82.8\%       & 42.8\% \\
gpt-oss-20b  & 80\% & & 92.9\% & 7.1\%  & 4.8\%  & & 83.1\% & 5.1\%  & 11.9\% & & 79.6\%       & 37.9\% \\
\bottomrule
\end{tabular}
\end{table*}

\begin{table*}[t]
\centering\small
\caption{Results from revised SV-COMP experiments on \texttt{valid-deref}
($N{=}10$ trials per phase), using the refined prompts; the answer key needed
no revision. ``CC'' is the consistency-check threshold (``no'' = no consistency check);
The \emph{Precision} is provided at the benchmark prevalence ($\pi{=}42\%$)
and at $\pi{=}10\%$.}
\label{tab:svcomp-v2}
\renewcommand{\arraystretch}{1.1}
\begin{tabular}{l c c p{0.5em} ccc p{0.5em} ccc p{0em} cc}
\toprule
& & & & \multicolumn{3}{c}{Actually true (42)} & & \multicolumn{3}{c}{Actually false (59)} & & \multicolumn{2}{c}{Precision} \\
\cmidrule(lr){5-7}\cmidrule(lr){9-11}\cmidrule(lr){13-14}
LLM          & LRE & CC   & & Recall  & Wrong & Uncert & & Spec.   & Wrong  & Uncert & & $\pi{=}42\%$ & $\pi{=}10\%$ \\
\midrule
o4-mini      & no  & no   & & 100.0\% & 0.0\% & 0.0\%  & & 99.8\%  & 0.2\%  & 0.0\%  & & 99.8\%       & 98.5\% \\
o4-mini      & no  & maj  & & 100.0\% & 0.0\% & 0.0\%  & & 100.0\% & 0.0\%  & 0.0\%  & & 100.0\%      & 100.0\% \\
o4-mini      & no  & 80\% & & 100.0\% & 0.0\% & 0.0\%  & & 100.0\% & 0.0\%  & 0.0\%  & & 100.0\%      & 100.0\% \\
o4-mini      & yes & 80\% & & 100.0\% & 0.0\% & 0.0\%  & & 100.0\% & 0.0\%  & 0.0\%  & & 100.0\%      & 100.0\% \\
\addlinespace
gpt-oss-120b & no  & no   & & 99.8\%  & 0.2\% & 0.0\%  & & 98.8\%  & 1.2\%  & 0.0\%  & & 98.4\%       & 90.3\% \\
gpt-oss-120b & no  & maj  & & 100.0\% & 0.0\% & 0.0\%  & & 100.0\% & 0.0\%  & 0.0\%  & & 100.0\%      & 100.0\% \\
gpt-oss-120b & no  & 80\% & & 100.0\% & 0.0\% & 0.0\%  & & 98.3\%  & 0.0\%  & 1.7\%  & & 97.7\%       & 86.8\% \\
gpt-oss-120b & yes & 80\% & & 100.0\% & 0.0\% & 0.0\%  & & 100.0\% & 0.0\%  & 0.0\%  & & 100.0\%      & 100.0\% \\
\addlinespace
gpt-oss-20b  & no  & no   & & 98.6\%  & 1.4\% & 0.0\%  & & 96.6\%  & 3.4\%  & 0.0\%  & & 95.5\%       & 76.4\% \\
gpt-oss-20b  & no  & maj  & & 100.0\% & 0.0\% & 0.0\%  & & 96.6\%  & 3.4\%  & 0.0\%  & & 95.5\%       & 76.6\% \\
gpt-oss-20b  & no  & 80\% & & 100.0\% & 0.0\% & 0.0\%  & & 96.6\%  & 0.0\%  & 3.4\%  & & 95.5\%       & 76.6\% \\
gpt-oss-20b  & yes & 80\% & & 100.0\% & 0.0\% & 0.0\%  & & 100.0\% & 0.0\%  & 0.0\%  & & 100.0\%      & 100.0\% \\
\addlinespace
gpt-4o       & no  & no   & & 93.6\%  & 6.4\% & 0.0\%  & & 89.3\%  & 10.7\% & 0.0\%  & & 86.4\%       & 49.3\% \\
gpt-4o       & no  & maj  & & 97.6\%  & 2.4\% & 0.0\%  & & 91.5\%  & 8.5\%  & 0.0\%  & & 89.3\%       & 56.1\% \\
gpt-4o       & no  & 80\% & & 97.6\%  & 2.4\% & 4.8\%  & & 84.7\%  & 3.4\%  & 11.9\% & & 82.3\%       & 41.6\% \\
gpt-4o       & yes & 80\% & & 100.0\% & 0.0\% & 4.8\%  & & 89.8\%  & 8.5\%  & 1.7\%  & & 87.7\%       & 52.2\% \\
\bottomrule
\end{tabular}
\end{table*}

\paragraph{Revised results.} 
SV-COMP's answer key needed no correction, but our initial prompt did not match two of
the suite's conventions. For the revised round we added two rules:
assume that \texttt{malloc}/\texttt{calloc}/\texttt{realloc} never fail,
and flag the use of a stack-allocated variable-length array (VLA) where the
array size isn't reasonably limited.
As shown in Table~\ref{tab:svcomp-v2}, these changes lift
the three reasoning LLMs to perfect recall (with the consistency check),
and \ttt{o4-mini} also achieves 100\% specificity.

\subsection{LLM reasoning evaluation (LRE)}
\label{sec:lre}
Sec.~\ref{sec:approach} introduced LRE, in which the LLM reconciles its own
discordant runs by weighing their reasoning.
In the revised FormAI experiments (Table~\ref{tab:formai-v2}), with the
non-frontier models, using LRE+CC
instead of only CC reduces the \ttt{uncertain} rate and improves
specificity.  The frontier models are already at or near the ceiling of 100\%
(for both recall and specificity) with CC alone,
so LRE adds little there.  For \texttt{gpt-oss-20b}, LRE+CC cuts uncertainty
on actually-true alerts from 29.8\% to 7.1\% and on actually-false alerts from
7.0\% to 1.9\%, raising specificity from 92.2\% to 97.0\%.  Recall remains
high, although sometimes suffers a little (e.g., a decrease in recall from
99.6\% to 99.4\% for \texttt{gpt-oss-20b}).

\paragraph{Comparison to plain majority voting.}
The \qt{\ttt{CC=maj}} rows of Table~\ref{tab:formai-v2} compare LRE and the
consistency check against the plain majority-vote baseline,
which is a natural alternative both to LRE and to CC.
The same phase-1 trials were used for all configurations.
We report the plain-majority result for the mid-tier reasoning LLMs and
for gpt-4o; we omit it for the frontier models (gpt-5.4, gpt-5.5, opus-4.8),
whose single-trial accuracy is already near-perfect, leaving little headroom
for the voting scheme to matter.
Majority voting is itself a modest improvement over a single trial (the
\qt{CC=no} rows): for gpt-oss-120b it lifts recall from 98.3\% to 99.4\% and
specificity from 94.4\% to 97.4\%.
With a suitable threshold, LRE+CC matches or exceeds majority voting on both
recall and specificity for all three mid-tier reasoning LLMs.
Using the 70\% threshold, o4-mini reaches
99.4\%~recall / 96.7\%~specificity versus majority voting's
96.9\% / 93.0\%;
gpt-oss-20b reaches 99.0\% / 97.4\% versus 98.4\% / 93.0\%; and
gpt-oss-120b reaches 100.0\% / 97.8\% versus 99.4\% / 97.4\%.
At the 80\% threshold, gpt-oss-120b instead trades a small amount of
specificity for recall (100.0\% / 96.3\% versus 99.4\% / 97.4\%).

The consistency check alone (especially at \ttt{CC=80\%}) tends to trades some specificity for
recall.  Marking difficult-to-adjudicate alerts as ``\ttt{uncertain}'' also
informs the analyst and/or developer that the alert
involves tricky reasoning about the program and/or ambiguities about what
exactly constitutes a real flaw.
With the \qt{\ttt{-{-}explain-uncertain}} option, LASAA makes an additional
query to the LLM to the summarize the key points of disagreement among
the LLM's replies to the original query.

The one place majority voting is competitive is the non-reasoning gpt-4o, where
it yields the highest specificity in that block (96.3\%); but even there it
confuses \ttt{true} and \ttt{dependent} on 34.7\% of positive alerts (the
\emph{DepMix} column), far more than with CC=80\% with or without LRE.

\subsection{Trigger test}
We applied the trigger test to 76 FormAI alerts whose ground truth we had
manually adjudicated (30 true positives, 19 dependent, 27 false positives).
These are the revised (v2) manual adjudications used in the FormAI experiment
(Sec.~\ref{sec:eval}).
Table~\ref{tab:trigger} cross-tabulates, for each ground-truth class, whether
the test produced a \emph{valid} driver (triggered the alert at the flagged line and
passed the validity check), an \emph{invalid} driver
(was rejected by the validity check), or a \emph{failed} driver
(failed to trigger the alert or no driver at all).

\begin{table}[t]
\centering\small
\caption{Trigger-test outcomes vs.\ manual ground truth (FormAI, 76 alerts).}
\label{tab:trigger}
\begin{tabular}{lrrrr}
\toprule
Ground Truth & Valid & Invalid & Failed & Total \\
\midrule
True       & 18 & 8  & 4  & 30 \\
Dependent  & 2  & 3  & 14 & 19 \\
False      & 0  & 9  & 18 & 27 \\
\midrule
Total      & 20 & 20 & 36 & 76 \\
\bottomrule
\end{tabular}
\end{table}

Two results stand out: First, the validity check helped to make the test trustworthy:
of the 40 drivers that fired, it rejected all 9 that targeted
false positives and 3 of the 5 that targeted dependent alerts, so \emph{no}
false positive yielded a valid trigger. Of the 20 valid triggers, 18
correspond to true positives and only 2 to dependent alerts (both borderline,
discussed below); a valid trigger is therefore strong evidence of a real
flaw. Second, the test is incomplete on the positive side: it
confirms 18 of 30 true positives (60\%) with a valid trigger, while 8 true
positives fired only via precondition-violating drivers (correctly discarded)
and 4 failed to fire at all.

Overall, the trigger outcome agrees with the manual verdict on 50 of the 76
alerts (66\%)---a valid trigger for a true positive, or no trigger for a
false or dependent alert---disagrees on 6 (7.9\%), and is inconclusive (a
driver fired but was rejected as invalid) on the remaining 20 (26\%).

\paragraph{Why some true positives go unconfirmed.}
The 4 true positives that never fired expose limits of the dynamic trigger test
rather than of the adjudication.
In these cases the driver under- or mis-targets the flaw (e.g., overflowing a buffer by too
little to fault, or passing a benign argument), fails to build, or trips
earlier undefined behavior that invalidates the trigger test. The two dependent alerts that
produced \emph{valid} triggers are cases the validity check should ideally
have rejected: One driver reaches the flagged line only by smuggling a
\texttt{NULL} pointer inside a struct field rather than as a direct argument,
and the other relies on a sign misinterpretation introduced on the line just
before the flagged one.

\newcolumntype{Y}{>{\raggedright\arraybackslash}X}

\def\pCE{p_{\itc{CE}}}
\def\pCU{p_{\itc{CU}}}
\def\ple{p_{1\trm{e}}}
\def\pz{\phantom{0}}
\begin{table*}[t]
\centering
\small
\begin{tabularx}{\textwidth}{@{}lYYY@{}}
\toprule
$t/N$ (pct) & $\phantom{\pCE=M}\ple=75\%$ & $\phantom{\pCE=M}\ple=50\%$ & $\phantom{\pCE=M}\ple=25\%$ \\
\midrule
$8/10$ (80\%) &
$\pCE=52.6\%,\ \pCU=47.4\%$     & $\pCE=\pz{}5.5\%,\ \pCU=89.1\%$ &  $\pCE=0.04\%,\ \pCU=47.4\%$ \\

$7/10$ (70\%)  &
$\pCE=77.6\%,\ \pCU=22.1\%$     &  $\pCE=17.2\%,\ \pCU=65.6\%$    &  $\pCE=0.4\%,\ \ \pCU=22.1\%$ \\ \addlinespace

$6/8$ \pz(75\%)  &
$\pCE=67.9\%,\ \pCU=31.7\%$     &  $\pCE=14.5\%,\ \pCU=71.1\%$    &  $\pCE=0.4\%,\ \ \pCU=31.7\%$ \\

$5/8$ \pz(62\%)  &
$\pCE=88.6\%,\ \pCU=\pz{}8.7\%$ &  $\pCE=36.3\%,\ \pCU=27.3\%$    &  $\pCE=2.7\%,\ \ \pCU=\pz{}8.7\%$ \\ \addlinespace

$5/6$ \pz(83\%)  &
$\pCE=53.4\%,\ \pCU=46.1\%$     &  $\pCE=10.9\%,\ \pCU=78.1\%$    &  $\pCE=0.5\%,\ \ \pCU=46.1\%$ \\

$4/5$ \pz(80\%)  &
$\pCE=63.3\%,\ \pCU=35.2\%$     &  $\pCE=18.8\%,\ \pCU=62.5\%$    &  $\pCE=1.6\%,\ \ \pCU=35.2\%$ \\

$3/4$ \pz(75\%)  &
$\pCE=73.8\%,\ \pCU=21.1\%$     &  $\pCE=31.3\%,\ \pCU=37.5\%$    &  $\pCE=5.1\%,\ \ \pCU=21.1\%$ \\

$2/2$ (100\%)  &
$\pCE=56.2\%,\ \pCU=37.5\%$     &  $\pCE=25.0\%,\ \pCU=50.0\%$    &  $\pCE=6.2\%,\ \ \pCU=37.5\%$ \\
\bottomrule
\end{tabularx}
\caption{Probability that the consistency check produces an incorrect adjudication ($\pCE$) or an adjudication of ``uncertain'' ($\pCU$), for different thresholds and single-trial error probabilities ($\ple$).  Note: numbers were rounded individually, so percentages expected to sum to 100\% (e.g., 77.6\% + 22.1\% + 0.4\% in the 7/10 row) might not sum to 100\% exactly.}
\label{tab:cc-error-uncertain}
\end{table*}

\begin{table}[t]
\centering\small
\renewcommand{\arraystretch}{1.1}
\begin{tabular}{@{}rcccccc@{}}
\toprule
$\ple$ & $t/N$ & (pct) & Recall & Spec. & $\pCE$ & $\pCU$ \\
\midrule
5\%  & 2/3   &  66\% & 99.28\% & 99.28\% & 0.73\% & 0.00\% \\
10\% & 3/5   &  60\% & 99.14\% & 99.14\% & 0.86\% & 0.00\% \\
15\% & 4/6   &  66\% & 99.41\% & 95.27\% & 0.59\% & 4.15\% \\
20\% & 5/9   &  55\% & 98.04\% & 98.04\% & 1.96\% & 0.00\% \\[1.0ex]
25\% & 8/14  &  57\% & 98.97\% & 96.17\% & 1.03\% & 2.80\% \\
30\% & 11/20 &  55\% & 98.29\% & 95.20\% & 1.71\% & 3.08\% \\
35\% & 20/38 &  52\% & 98.07\% & 95.93\% & 1.93\% & 2.14\% \\
40\% & 44/85 &  51\% & 98.15\% & 95.06\% & 1.85\% & 3.09\% \\
\bottomrule
\end{tabular}
\caption{For each single-trial error rate ($\ple$), the smallest number of
trials $N$ (with an accompanying threshold $t$) for which the consistency check
attains recall $\ge 98\%$ and specificity $\ge 95\%$.
Recall${}=1-\pCE$ and specificity${}=1-\pCE-\pCU$, treating ``uncertain''
as positive.
}
\label{tab:cc-min-trials}
\end{table}

\begin{table}[ht]
\centering\small
\renewcommand{\arraystretch}{1.2}
\begin{tabular}{r c c}
\toprule
            & Count of actually & Count of actually \\
Error rate  & false alerts      & true/dep alerts   \\
\midrule
0\%        & 10 & 32 \\
1\%        & 7  & 4  \\
2\%        & 3  & 6  \\
3\%--7\%   & 4  & 1  \\
13\%--16\% & 1  & 5  \\
59\%       & 1  & 0  \\
65\%--66\% & 1  & 1  \\
\midrule
TOTAL      & 27 & 49 \\
\bottomrule
\end{tabular}
\caption{Observed error rates for \ttt{gpt-oss-20b} on individual FormAI alerts}
\label{tab:error-rate-gpt-oss-20b}
\end{table}

\begin{table}[t]
\centering\small
\begin{tabular}{crccc}
\toprule
$t/N$ & (pct) &  Recall  &   Spec. \\
\midrule
  1/1 &  100\% &  96.88\% &   93.56\% \\[1ex]

  2/2 &  100\% &  98.91\% &   90.10\% \\[1ex]

  2/3 &  66\% &   97.97\% &   94.63\% \\
  3/3 &  100\% &  99.39\% &   87.83\% \\[1ex]

  3/4 &  75\% &   98.72\% &   93.08\% \\
  4/4 &  100\% &  99.61\% &   86.08\% \\[1ex]

  3/5 &  60\% &   98.20\% &   94.59\% \\
  4/5 &  80\% &   99.07\% &   92.08\% \\
  5/5 &  100\% &  99.74\% &   84.58\% \\[1ex]

  4/6 &  67\% &   98.60\% &   93.54\% \\
  5/6 &  83\% &   99.31\% &   91.35\% \\
  6/6 &  100\% &  99.83\% &   83.23\% \\[1.5ex]

  5/8 &  62\% &   98.50\% &   93.61\% \\
  6/8 &  75\% &   99.08\% &   92.42\% \\[1.5ex]

  7/10 & 70\% &   98.89\% &   92.74\% \\
  8/10 & 80\% &   99.42\% &   91.80\% \\[1.5ex]

 74/99 & 75\% &   99.92\% &   92.59\% \\
 78/99 & 79\% &   99.99\% &   92.53\% \\
\bottomrule
\end{tabular}
\caption{Simulated recall and specificity for \ttt{gpt-oss-20b} on FormAI for
various values of number of trials $N$ and threshold $t$, using the observed
error rate for each alert.}
\label{tab:sim-formai-20b}
\end{table}

\section{Using LASAA in Practice}
\label{sec:practice}

Our evaluation suggests several concrete recommendations for deploying LASAA on
a real alert backlog.

\paragraph{Triage.}
LASAA is best positioned as a triage filter: its operational value is to
convert the large pool of unreviewed alerts into a much smaller pool that
analysts and/or developers actually look at.
LASAA discards only the alerts it adjudicates \ttt{false}. Every \ttt{true},
\ttt{dependent}, or \ttt{uncertain} alert is returned for human attention.
This preserves the property that matters in high-assurance settings: a real
flaw should not be silently cleared.
Analysts/developers then spend their limited attention on the returned alerts
--- and, for \ttt{dependent} alerts, on the single upstream line that LASAA
identifies as the one to fix, rather than on every downstream symptom.

\paragraph{Tune the consistency-check settings to trade off FPs, FNs, and token budget.}
The two types of adjudication errors have very different costs: a missed flaw can
field a real vulnerability, whereas a false alarm merely adds review burden.
The consistency-check settings (threshold and number of trials) are the main
mechanism for managing missed flaws vs.\ excessive false alarms.
The number of trials ($N$) also directly affects
the token cost of adjudicating each alert.
To ensure that the consistency check (CC) is well-defined, the threshold number
of trials ($t$) must be strictly greater than $N/2$.  Note that if $N$ is odd
and $t = N/2 + 1/2$, then CC
can never return a verdict of ``\ttt{uncertain}'' (unless there is a three-way
split between true, false, and dependent).

For the situation where the LLM can return only ``true'' or ``false''
(not ``dependent'' or ``uncertain''),
Table~\ref{tab:cc-error-uncertain} shows, for various thresholds and number of trials,
the probability of CC returning an erroneous verdict ($\pCE$) and the
probability of CC returning a verdict of ``\ttt{uncertain}'' ($\pCU$),
assuming that the probability of a single trial returning an erroneous verdict
($\ple$) is 25\%, 50\%, or 75\%.
In this table,
we calculated $\pCE$ and $\pCU$ by modeling the number of erroneous
single-trial adjudications as a binomial random variable over $N$ independent
trials, where each trial has probability $\ple$ of being erroneous.

Note that, for a given $t$ and $N$, the value of $\pCE$ at $\ple{=}75\%$ (i.e., the
probability that CC returns the \itc{wrong} adjudication when $\ple{=}75\%$) is equal
to the probability that CC returns the \itc{right} adjudication at $\ple{=}25\%$.
The reason is that these two scenarios are mirror images: a trial is erroneous
with probability $75\%$ in the first scenario, just like a trial is correct with
probability $75\%$ in the second. So the number of erroneous trials in the
first scenario has the same distribution as the number of correct trials in the
second, and CC errs (needs $\ge t$ erroneous trials) in the first exactly as
often as it succeeds (needs $\ge t$ correct trials) in the second.

The optimal settings depend on the code being analyzed, the static-analysis
tools that produce the alerts, and the LLM that is used to answer queries.
When these change significantly in a relevant way, it would be desirable
to have an automated way of optimizing the settings; this is left for
future work.

Let us consider a simple model where the LLM has the same error rate for every
alert.  With this model, Table~\ref{tab:cc-min-trials} reports, for a range of
single-trial error rates, the smallest $N$ (and an accompanying $t$) for which
the consistency check attains recall $\ge 98\%$ and specificity $\ge 95\%$ in
the binary-verdict case, assuming a very large number of alerts.
The required $N$ rises sharply as the single-trial error rate grows, so a more
error-prone LLM/workload costs proportionally more tokens to adjudicate at the
same accuracy.

In practice, unlike the simple model discussed above, the single-trial error
rate $\ple$ often varies considerably from alert to alert.
For example, \ttt{gpt-oss-20b}'s per-alert error rates on FormAI (measured by
querying the model 100 times per alert and counting how often
the single-trial verdict was wrong) range from 0\% to 66\%, with the
distribution shown in Table~\ref{tab:error-rate-gpt-oss-20b}.
To see how the consistency check is affected by the choice of threshold and
number of trials under this more realistic setting,
we simulated it directly from these per-alert rates.
For each alert $i$ with observed single-trial error rate $\ple^{(i)}$ and each
candidate setting $(N,t)$, we computed $\pCE^{(i)}$ and $\pCU^{(i)}$ by
plugging $\ple^{(i)}$ into the same binomial calculation used for
Table~\ref{tab:cc-min-trials}.
We then aggregated across alerts, labeling each alert positive if its ground truth
is \ttt{true} or \ttt{dependent} and negative if its ground truth is
\ttt{false}:
\begin{eqnarray*}
\mathrm{Recall} & = & \operatorname*{mean}_{i \in \mathrm{positives}} \bigl(1-\pCE^{(i)}\bigr),
\\
\mathrm{Specificity} & = & \operatorname*{mean}_{i \in \mathrm{negatives}} \bigl(1-\pCE^{(i)}-\pCU^{(i)}\bigr).
\end{eqnarray*}
Table~\ref{tab:sim-formai-20b} reports the resulting recall and specificity
for several $(N,t)$ settings.
Note that these are the \itc{expected values}; actual experiments will show
variation, generally with greater variation at smaller $N$.
Since 2 of the 27 actually false alerts have an error rate greater than 50\%,
the maximum specificity approaches \hbox{25/27 = 92.59\%} as $N$ increases.

\paragraph{Choosing a model.}
Use a reasoning LLM. In our experiments the larger but non-reasoning
\texttt{gpt-4o} did noticeably worse than even the small reasoning model
\ttt{gpt-oss-20b}.
Among reasoning models, the choice is driven by data-sensitivity constraints,
available hardware, and budget.
Where source code cannot leave the premises, the open-weight \texttt{gpt-oss}
models run locally with no data egress; \texttt{gpt-oss-20b} is small enough to
be run on relatively inexpensive hardware yet still performed quite well when
paired with LRE and a consistency check. Where sending code to a hosted API is
acceptable, frontier closed models give the highest accuracy.
Whichever model is chosen, do not over-trust accuracy measured on
benchmark-style code: our filename-reconstruction probe (Sec.~\ref{sec:eval})
shows substantial Juliet memorization, so a model's numbers on familiar suites
may overstate its performance on unfamiliar production code.

\paragraph{Add LRE to control review burden.}
The consistency check often increases recall at the cost of a higher \ttt{uncertain}
rate, and every \ttt{uncertain} verdict is an alert a human must still review.
Adding the LRE step on top of the consistency check sharply reduces that
\ttt{uncertain} rate for the non-frontier models and, in our FormAI
experiments, improves their specificity.  The remaining wrong-answer rates are
still low, but not always lower than in the consistency-check-only rows.

LRE roughly doubles the token cost of alerts to which it is applied (assuming
that input tokens cost much less than output tokens, as is usual),
but it is applied only to alerts where the replies to the original query
aren't unanimous.  On the FormAI benchmark, LRE was applied to
19\% of alerts for \ttt{o4-mini},
26\% of alerts for \ttt{gpt-oss-120b}, and
43\% of alerts for \ttt{gpt-oss-20b}.

\paragraph{Use the trigger test to prioritize confirmed bugs.}
For alerts adjudicated \ttt{true}, the dynamic trigger test
(Sec.~\ref{sec:trigger}) supplies independent, execution-based evidence.
In our evaluation it never produced a \emph{valid} trigger for a false
positive, so a valid trigger is strong evidence of a genuine flaw; such alerts
can be moved to the front of the repair queue with high confidence.
However, the trigger test confirmed only 60\% of true positives, so
the \emph{absence} of a trigger is only weak evidence of exoneration.
The trigger test also needs code that compiles and runs under sanitizers, so it
is most useful on buildable subsystems rather than on isolated source code snippets.

\paragraph{Read the reasoning, not just the verdict.}
Because every verdict comes with a justification, a reviewer can audit
\emph{why} the LLM reached a conclusion rather than trusting the label.
In our own work, the justifications repeatedly exposed mistakes in our manual
answer key for FormAI (Sec.~\ref{sec:groundtruth}), illustrating the
high quality of the LLM's explanation of its verdict (in particular,
high quality-of-thinking, as opposed to high quality-of-writing).
For \ttt{uncertain} verdicts, LASAA can summarize the competing arguments made
by different runs of the LLM and pinpoint the sources of disagreement,
easing the work of an analyst reviewing the alert.

\paragraph{Diagnose systematic errors before distrusting the model.}
When many similar alerts are adjudicated incorrectly by the LLM, read the reasoning
before concluding that the model is incapable. In our experiments, many apparent
errors were not hallucinations or faulty reasoning but a mismatch between how the
LLM and the answer key interpreted an ambiguous question. This can often be fixed
by adding clarifications to the prompt so that the LLM's interpretation matches
the intended convention. Because every verdict comes with its reasoning,
these cases are easy to recognize: the justification is internally sound and
simply answers a slightly different question.

\section{Related Work}
\label{sec:related}

\subsubsection*{LLM-based adjudication}
Li et al.\ showed GPT-4 can adjudicate use-before-initialization bugs in the Linux
kernel~\cite{li2023assisting,li2023hitchhiker,li2024enhancing}; the follow-on
\emph{BugLens} introduces a structured post-refinement workflow that sharply
improves precision on kernel taint bugs~\cite{li2025hitchhiker2}.

ZeroFalse~\cite{iranmanesh2025zerofalse} enriches CodeQL/SARIF alerts with
data-flow traces and CWE-specific rubrics, and then asks an LLM to adjudicate.

Wagner et al.~\cite{wagner2025complementary}, like LASAA, use LLMs to
adjudicate static-analysis alerts.  They find that false-positive detection is
improved by both (1) explicitly prompting for chain-of-thought (on older pre-o1
LLMs that don't automatically reason in a hidden block before producing
output) and (2) the self-consistency approach of Wang~\cite{wang2022self}.
Note that Wang's self-consistency approach differs from LASAA's consistency check
in that Wang's always returns the most common final answer, whereas LASAA's
returns ``uncertain'' if the specified consistency threshold is not met.

Du et al.~\cite{du2026tencent} study LLM-based false-positive reduction in an
industrial setting at Tencent, using 433 alerts from Tencent's proprietary
BkCheck analyzer covering null-pointer dereference, out-of-bounds, and
divide-by-zero warnings.
They evaluate four LLMs (GPT-4o, Claude-Opus-4, Qwen-3-Coder, and DeepSeek-R1)
under a variety of techniques, including LLM4SA~\cite{wen2024llm4sa} and
LLM4PFA~\cite{du2025pfa} (both described below).
In their results, LLM4PFA is the most effective technique, eliminating 94--98\%
of false positives while achieving true-bug recall of 0.75--0.88 and an F1
score of 0.83--0.86.
Both this study and LLM4PFA measure false-positive filtering with a metric they
call \emph{False Positive Reduction Recall}:
$\mathrm{FPR\_R} = \mathrm{TN}/(\mathrm{TN}+\mathrm{FP})$, the recall over the
negative class.  FPR\_R is identical to what we call \emph{specificity}.

\subsubsection*{Engineering the code context and reasoning}
A related line of work treats the LLM as an expert reviewer and concentrates on
what to feed it; the three systems below form a clear progression.
LLM4SA~\cite{wen2024llm4sa} is the broad baseline: it normalizes warnings from
several analyzers, traverses program-dependence graphs to extract a relevant
code slice per alert, and prompts an LLM to classify each as a real bug or a
false alarm.
LLM4FPM~\cite{chen2024llm4fpm} keeps that pipeline but argues its bottleneck is
context \emph{quality}: it builds an extended code property graph, slices
warning-relevant lines, and
identifies additional dependencies for enriched cross-file context.
LLM4PFA~\cite{du2025pfa} instead restructures the \emph{reasoning} for a
narrower class of warnings---rather than judge a whole call chain at once, it
decomposes source-to-sink reachability into constraint extraction, LLM-assisted
range reasoning, and Z3 solving, specializing in path feasibility analysis (PFA).
LASAA differs from all three: it retrieves context through an interactive loop
driven by the LLM's own \texttt{need\_defs} requests rather than a fixed
slicing pipeline.

\subsubsection*{LLMs for tasks other than adjudication}
IRIS~\cite{li2024iris} uses an LLM to
infer taint specifications and perform contextual analysis for
whole-repository detection, finding more vulnerabilities than CodeQL on
CWE-Bench-Java.

CodeCureAgent~\cite{joos2025codecureagent} goes beyond adjudication to
automatically repair true positives in Java projects.

Wu et al.~\cite{wu2023lemur} report success in using LLMs for generating
formal-verification proofs, beating state-of-the-art formal-verification tools
on a number of hard cases.

\section{Limitations and Future Work}
\label{sec:remaining}

\paragraph{Threats to validity.}
Benchmark contamination (for Juliet and likely for SV-COMP too) is real and, per our filename-reconstruction
probe for Juliet (Sec.~\ref{sec:eval}), higher than we had initially assessed: o4-mini can
often recover the identity of a sanitized Juliet case, so the Juliet numbers
cannot be read as strong evidence of generalization. We mitigate this by basing our
generalization claims primarily on FormAI. Our flipped-verdict experiment
(Sec.~\ref{sec:eval}) provides complementary evidence that, even on Juliet,
o4-mini analyzes the code rather than reciting memorized verdicts, but it covers
only 20 cases and does not by itself establish generalization to real-world code.
Finally, Juliet, FormAI, and SV-COMP are not representative of real-world code;
evaluating LASAA on real-world repositories at scale remains future work.

\paragraph{No static analyzer in the loop.}
Although LASAA is built to ingest alerts from static analyzers, the alerts in
our three evaluations were not produced by one.  Alerts from actual static
analyzers usually have at least four parts: filename, line number,
alert category (e.g., CWE number), and a message describing the flaw (perhaps
just the title of the CWE number, or perhaps more detailed).
For Juliet, the LLM was asked if a given CWE occurred \itc{anywhere} in the
source code, rather than on a particular line.
Likewise, for SV-COMP, the LLM was asked if a memory-safety violation (e.g.,
buffer overflow, use-after-free, dereferencing an uninitialized pointer)
occurred \itc{anywhere} in the source code.
In general, determining whether a particular type of flaw occurs anywhere is
harder than determining whether such a flaw occurs on a given line,
so we do not expect the lack of a specified line number to paint a rosier
picture than we would get if a line number were specified.
For FormAI, the LLM is given a line number, but the alert categories and alert
message are different from those typically given by actual static analyzers.
Additionally, alerts from real static analyzers may differ in weakness-type mix
and the kinds of code constructs that provoke false positives.
Our results therefore do not measure how well the pipeline performs on the
output of any particular static-analysis tool; evaluating LASAA on alerts
produced by widely used static analyzers is part of the planned future work on
real-world codebases.

\paragraph{Difficulty and ambiguity of ground truth.}
Establishing ground truth was one of the hardest parts of this study
(Sec.~\ref{sec:groundtruth}), and it is itself a threat to validity.
Some alerts are genuinely ambiguous rather than cleanly true or false. Where an
answer key encodes difficult judgment calls, reported accuracy is measured
against one defensible labeling, and a different careful adjudicator might
score the same verdicts differently.

\paragraph{Sample sizes and generalization.}
Our samples are small relative to the suites they are drawn from: 100 Juliet
cases (out of 60{,}000+) and 76 FormAI alerts (out of 331{,}000), of which only
27 are actually false. Two distinct sources of uncertainty are worth separating.
For the \emph{specific} alerts we sampled, running many trials per alert averages
out the LLM's nondeterminism, so we can estimate recall and specificity \emph{on
that subset} reasonably precisely. Using pooled 99\% exact intervals on the
no-mitigation rows (\ttt{LRE=no}, \ttt{CC=no}) for the reasoning models tested
on all three suites in the revised experiments, the maximum amount by which the 99\% confidence intervals
extend above/below the point estimate are,
by benchmark: for Juliet, +1.07/-1.54 percentage points for recall and
+1.28/-1.74 for specificity; for FormAI, +0.63/-0.72 for recall and
+1.16/-1.32 for specificity; and for SV-COMP, +1.06/-2.26 for recall and
+1.62/-2.41 for specificity.
For the FormAI LRE+CC rows for those same models, where we have 10
final-adjudication replications, the corresponding maxima are 0.80 above / 1.84
below for recall and 2.85 above / 4.52 below for specificity.
The single-replication Juliet and SV-COMP CC rows have wider final-adjudication
intervals; for example, 42/42 correct recall on SV-COMP has a 99\% confidence
interval whose lower endpoint is 88.15\%.

Extrapolating these rates to the \emph{whole} benchmark is a separate question,
governed by the number of alerts sampled rather than the number of trials: a
rate computed over roughly 100 alerts (and, for FormAI specificity, over just
27 actually-false alerts) carries a confidence interval too wide to support
strong claims about whole-benchmark performance.

\paragraph{Adaptive number of trials.}
Currently, the consistency check runs a fixed number of trials for each alert.
A better approach would be to first run $N_1$ trials of an alert and then run
an additional $N_2$ trials iff the results of the first $N_1$ trials are not
unanimous.

\paragraph{Automated tuning.}
LASAA has a few settings (e.g., threshold and number of trials) that can be
tuned.  Different settings might be optimal for different combinations of
workload and LLM.
For example, if the single-trial error rate for most alerts is around 30\%,
then more trials and/or a higher threshold would be needed to achieve
satisfactory recall and specificity than if the error rate is around 10\%.
(See Table~\ref{tab:cc-min-trials}: at a constant 10\%
single-trial error rate, $N{=}5$ trials suffice for recall $\ge 98\%$ and
specificity $\ge 95\%$, whereas at a constant error rate of 30\%, it takes $N{=}20$.)
Investigating automated ways of tuning these parameters (either completely
automated or aided by a number of manual adjudications) is future work.


\section{Conclusion}
\label{sec:conclusion}

We set out to measure how well modern LLMs can adjudicate static-analysis
alerts. To that end, we built LASAA, an open, analyzer-agnostic tool that
adjudicates each alert with an LLM and reports a justification alongside every
verdict.  We ran LASAA on three benchmark suites (Juliet, FormAI, and SV-COMP)
to measure the LLM's performance.  Since LLMs are nondeterministic and
sometimes make mistakes, LASAA employs two mistake-mitigation methods: a
consistency check (CC) and an LLM reasoning evaluation (LRE).
In our evaluation on reasoning LLMs, LRE+CC matches or beats plain majority
voting with respect to both recall and specificity if a suitable threshold is
chosen.  Additionally, when the consistency check yields a verdict of
\ttt{uncertain}, the LLM is asked to summarize the key points of disagreement
between the LLM's answers, providing useful information for an analyst
subsequently adjudicating the alert.

Across these benchmark suites, the reasoning LLMs we tested on all three
suites (o4-mini, gpt-oss-120b, gpt-oss-20b) reach high recall and high
specificity, keeping low both types of adjudication error (missed flaws and
false alarms). On the revised experiments, each of these three LLMs reaches a
recall of at least 98\% and a specificity of at least 96\% --- on Juliet and
SV-COMP using the consistency check alone, and on FormAI using the consistency
check together with LRE --- with the single exception that o4-mini's
specificity on FormAI is 94.8\%.  The frontier reasoning models we tested on
FormAI (gpt-5.4, gpt-5.5, and opus-4.8) each reach 100\% recall and
100\% specificity there with the consistency check.

For independent, execution-based evidence, we developed a dynamic trigger tool,
which includes a validity check that rejects drivers that reach the flaw only
by violating the program's preconditions.  In our evaluation, the dynamic
trigger tool never produced a valid trigger for a false positive, indicating
that a valid trigger is strong evidence of a genuine flaw; however, it
confirmed only 60\% of the true positives, so the absence of a trigger is not
exoneration.

Taken together, the results indicate that modern reasoning LLMs, when paired
with mistake-mitigation methods, are highly accurate at static-analysis
adjudication, at least on the benchmark suites we tested.
The most important next step is to move beyond synthetic, LLM-generated, and
small hand-crafted test cases to large, real-world codebases.

\section*{Acknowledgments}
\small

Claude Opus 4.8 and GPT-5.5 were used to prepare initial drafts of text of this
paper; their output was carefully reviewed and revised by us.
Additionally, they (and earlier versions of them) were used to write some of
the source code in LASAA itself, including scripts for analyzing and reporting
results; again, all of their output was carefully reviewed and revised by us.

\section*{Document Markings}
\small
{

Copyright 2026 Carnegie Mellon University.

This material is based upon work supported by the Department of War under Air
Force Contract No. FA8702-15-D-0002 with Carnegie Mellon University for the
operation of the Software Engineering Institute, a federally funded research
and development center.

The opinions, findings, conclusions, and/or recommendations contained in this
material are those of the author(s) and should not be construed as an official
US Government position, policy, or decision, unless designated by other
documentation.

References herein to any specific entity, product, process, or service by trade
name, trademark, manufacturer, or otherwise, does not necessarily constitute or
imply its endorsement, recommendation, or favoring by Carnegie Mellon
University or its Software Engineering Institute nor of Carnegie Mellon
University - Software Engineering Institute by any such named or represented
entity.

\vspace{0.5ex}
\begin{spacing}{1.1}
NO WARRANTY. THIS CARNEGIE MELLON UNIVERSITY AND SOFTWARE ENGINEERING INSTITUTE
MATERIAL IS FURNISHED ON AN ``AS-IS'' BASIS. CARNEGIE MELLON UNIVERSITY MAKES
NO WARRANTIES OF ANY KIND, EITHER EXPRESSED OR IMPLIED, AS TO ANY MATTER
INCLUDING, BUT NOT LIMITED TO, WARRANTY OF FITNESS FOR PURPOSE OR
MERCHANTABILITY, EXCLUSIVITY, OR RESULTS OBTAINED FROM USE OF THE MATERIAL.
CARNEGIE MELLON UNIVERSITY DOES NOT MAKE ANY WARRANTY OF ANY KIND WITH RESPECT
TO FREEDOM FROM PATENT, TRADEMARK, OR COPYRIGHT INFRINGEMENT.
\end{spacing}

\vspace{0.5ex}
[DISTRIBUTION STATEMENT A] This material has been approved for public release
and unlimited distribution.  Please see Copyright notice for non-US Government
use and distribution.

This work is licensed under a Creative Commons Attribution-NonCommercial 4.0
International License (\url{https://creativecommons.org/licenses/by-nc/4.0/}).
Requests for permission for non-licensed uses should be directed to the
Software Engineering Institute at permission@sei.cmu.edu.

\vspace{0.5ex}
\noindent
This work product was created in part using generative AI.

\vspace{0.5ex}
\noindent
DM26-0698

}

\balance
\bibliographystyle{unsrturl}
\bibliography{lasaa}

\end{document}